\documentclass{article}

\usepackage{arxiv}

\usepackage[utf8]{inputenc} 
\usepackage[T1]{fontenc}    
\usepackage{hyperref}       
\usepackage{url}            
\usepackage{booktabs}       
\usepackage{amsfonts}       
\usepackage{nicefrac}       
\usepackage{microtype}      
\usepackage{lipsum}		
\usepackage{graphicx}
\usepackage{natbib}
\usepackage{doi}
\usepackage{
            amsmath,      
            amssymb,
            comment,
            pdfpages,
            color,
            colortbl,
            todonotes,
            etoolbox,       
            ifxetex,          
            ifluatex,         
            float,              
            xcolor,           
            color,             
            bm,                
            wrapfig,          
            }

\usepackage{listings}
\definecolor{codegreen}{rgb}{0,0.6,0}
\definecolor{codegray}{rgb}{0.5,0.5,0.5}
\definecolor{codepurple}{rgb}{0.58,0,0.82}
\definecolor{backcolour}{rgb}{0.95,0.95,0.92}

\usepackage{
            customstyle
            }

\title{Comparison of Deterministic \& Bayesian Calibration of MFiX-PIC, Part 1: Settling Bed}

\author{ \href{https://orcid.org/0000-0002-1661-2859}{\includegraphics[scale=0.06]{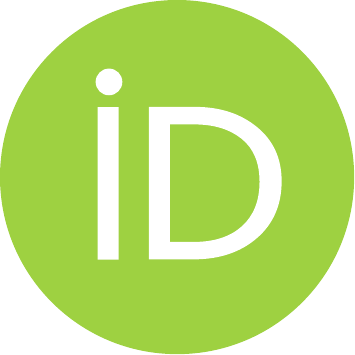}\hspace{1mm}Aytekin Gel $^1{^,}^2$}\thanks{Correspondance author, email: \texttt{aike@alpemi.com}} \\
    $^1$ National Energy Technology Laboratory,\\
    3610 Collins Ferry Road, \\
    Morgantown, WV 26505\\
	$^2$ ALPEMI Consulting, L.L.C.\\
	8205 S Priest Dr \#13951\\
	Tempe, AZ 85284 \\
	\And
	\href{https://orcid.org/0000-0002-1089-8163}{\includegraphics[scale=0.06]{orcid.pdf}\hspace{1mm}Avinash Vaidheeswaran $^1{^,}^3$} \\
    $^1$ National Energy Technology Laboratory,\\
    3610 Collins Ferry Road, \\
    Morgantown, WV 26505 \\
	$^3$ NETL Support Contractor,\\
	3610 Collins Ferry Road, \\
    Morgantown, WV 26505 \\
	\AND
    {Mary Ann Clarke $^1$} \\
    $^1$ National Energy Technology Laboratory, \\
    3610 Collins Ferry Road \\
    Morgantown, WV 26505 \\
}

\renewcommand{\shorttitle}{Comparison of Deterministic \& Bayesian Calibration of MFiX-PIC}

\hypersetup{
pdftitle={Comparison of Deterministic \& Bayesian Calibration of MFiX-PIC, Part 1: Settling Bed},
pdfsubject={q-bio.NC, q-bio.QM},
pdfauthor={Aytekin Gel, Avinash Vaidheeswaran, Mary Ann Clarke},
pdfkeywords={Granular Multiphase Flows, Gravitational Particle Settling, Computational Fluid Dynamics (CFD), Parcel-in-Cell~(PIC), Uncertainty Quantification, Deterministic Calibration, Bayesian Calibration},
}

\setcitestyle{yysep={;}}

\begin{document}
\maketitle

\vspace*{4mm}
\begin{abstract}
    Particle-in-Cell (PIC) approach for modeling dense granular flows has gained popularity in recent years due to its time to solution efficiency. The methodology is useful for modeling large-scale systems with a relatively lower computational cost. However, the method requires the definition of several empirical parameters whose effects are not well understood. A systematic approach to identify sensitivities and optimal settings of these parameters is required. Already, it is known that the choice of these parameters depends on a problem's flow regime.
    For instance, parameter values would be chosen differently for a settling bed or a fluidized bed. In this study, five different PIC model parameters were selected for calibration when applied to the case of particles settling in a dense medium. PIC implementation from the open-source software MFiX (MFiX-PIC) was used. This study extends the earlier work to assess the five model parameter settings using deterministic calibration  
    by employing a statistical calibration methodology commonly referred as Bayesian calibration. Results from deterministic calibration are compared with Bayesian calibration, and up to 6.5 fold improvement in prediction accuracy is observed with the latter approach. 
\end{abstract}

\section{Introduction}

Particle laden flows are common in many engineering applications including chemical, pharmaceutical, energy and food industries. Simulation-Based Engineering (SBE)
has widely been used to design, troubleshoot and optimize such systems while attempting to minimize their operational costs. Recently, there has been an increasing demand for modeling industrial-scale systems where application of conventional simulation techniques such as discrete element model (DEM) may be challenging. Tracking individual particles and their collisions with neighbors can become computationally intractable when particle count exceeds the order of tens of millions. Although there is rapid development in hardware resources for high-performance 
computing,
industrial scale models still may suffer from unreasonable computational turnaround times. Consequently, this issue has led to the development of coarse-graining strategies such as coarse grained DEM (CGDEM) 
or Particle-in-Cell (PIC), which present a trade-off between accuracy and time to solution. 

Recently, a systematic approach \citep{Gel2018} has been adopted by the Multiphase Flow Science group at the U.S. Department of Energy's National Energy Technology Laboratory (NETL) to generate high-quality data \citep{vaidheeswaran2017statistics,Gel2018,vaidheeswaran2020fluidization,vaidheeswaran2021chaos,vaidheeswaran2022validation,higham2020using,higham2021anomalous,gao2020comprehensive,lu2020experimental,li2022measuring}, with subsequent use for validation, sensitivity analysis, uncertainty quantification (UQ), or calibration. Several previous studies \citep{Vaidheeswaran2020,vaidheeswaran2021assessment} analyzed parametric sensitivities of different MFiX-PIC model parameters through global sensitivity analysis. Even though sensitivity analysis indicates the influence of selected parameters on quantities of interest (QoI) in a simulation, the method does not quantify ideal input parameter values. As such, the current effort explores deterministic calibration and Bayesian inference as a means to identify optimal single-value or ranged input parameters for particular MFiX-PIC simulations, respectively. In general, the model parameters are expected to depend on hydrodynamics as outlined in \cite{vaidheeswaran2021assessment}. The current study pertains to particles settling in a dense medium, where $U/U_{mf} < 1$. The work presented extends the prior study ~\citep{Gel2021TRS} where an optimal set of MFiX-PIC model parameters were identified. However, after the aforementioned work, MFiX-PIC developers corrected a software bug that may have influenced the original study. Hence, the simulations are re-run in this study, and the parameters obtained through deterministic calibration are updated. In addition, Bayesian inference is presented in the current study to provide distributions of model parameters.

The preprint 
is organized as follows: First, an overview of MFiX-PIC methodology is provided in Section 2. A background on calibration methodologies and softwares used is provided in Section 3. The problem setup is described in Section 4 followed by results from deterministic calibration and their comparison with results from Bayesian inference. The preprint also includes a separate data management and repository section (Appendix ~\ref{appendix:PS_RSM}).
which contains information necessary to replicate the calibration analyses.

\section{MFiX-PIC Overview}

\begin{figure}
    \begin{center}
    \includegraphics{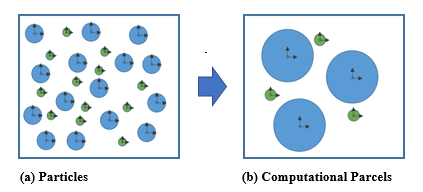}
    \end{center}
    \caption{Visual concept of poly-disperse particle consolidation to computational parcels \cite{PICTheoryClarke2020}. (a) A single
cell populated with particles. (b) The same single cell after a statistical weight has been applied to each solid phase. }
    \label{fig:ParticlesToParcels}
\end{figure}

\textcolor{black}
PIC methodology relies on representing particles as computational parcels. The number of particles in a parcel is determined by statistical weight, $W_{p}$, which is a user input. In this case study, $W_{p}$ is one of the model parameters considered for calibration. The individual particles inside each parcel are assumed to be spherical and have identical properties. In the case of systems having multiple components in the solids-phase, a separate statistical weight is assigned for each. The illustration in Figure~\ref{fig:ParticlesToParcels} shows a binary mixture, where $W_{p}$ = 5 for blue particles and $W_{p}$ = 4 for green particles.

The location of a parcel, $X_{i}$ is updated using,
\begin{equation}
    U_{i} = \frac{dX_{i}}{dt}
\end{equation}

\noindent where, the velocity $U_{i}$ is obtained by solving the linear momentum equation as follows \cite{PICTheoryClarke2020},
\begin{equation}
W_{p}m_{p}\frac{dU_{i}}{dt}=W_{p}(m_{p}g_{i}+
          \frac{m_{p}}{\epsilon_{s}\rho_{s}}
          \nabla \tau_{p}) + S_{mi}^{(p)}
\label{eq:mom}
\end{equation}

\noindent $g_{i}$ is the acceleration due to gravity acceleration in the coordinate direction $i$. $\rho_{s}$, $m_{p}$, $\epsilon_{s}$ represent density, mass, and volume fraction of solids-phase. $S_{mi}^{p}$ is the source term combining interphase drag and pressure gradient terms given by,
\begin{equation}
    S_{mi}^{p} = W_{p}(-\frac{dP_{g}}{dx_{i}}V^{(p)}+\beta_{g}^{(p)}V^{(p)}(U_{gi}-U_{i}^{(p)}))
\end{equation}

The implementation of an inter-particle stress term, $\tau_{p}$ in Eqn. ~\ref{eq:mom} is not straightforward. The stress is initially evaluated as,
\begin{equation}
\tau_{p} = \frac{P_{o} \epsilon_{s}^\beta}{\mathrm{max}[\epsilon_{s}^*-\epsilon_{s},\delta (1-\epsilon_{s})]}
\end{equation}

\noindent $P_{0}$, $\beta$, and $\epsilon_{s}^*$ are empirical parameters which are also considered for calibration in this study. The parameter $\delta$ is used to avoid singularities during computation and is set to $10^{-7}$ by default. Once $\tau_{p}$ is evaluated, its gradient is used to determine the actual contribution of inter-particle stress to parcel motion, $\delta u_{p}$, as explained by the following pseudo-algorithm \citep{vaidheeswaran2021assessment}:

\label{switchingalgo}

\hspace*{1cm} if  $\nabla\tau_{p}\leq$0 \\[2mm]
\hspace*{1.7cm}     PIC velocity contribution = min($\delta u_{p}$,$\alpha$*Slip Velocity) \\[1.5mm]
\hspace*{1.7cm}    PIC velocity contribution = max(PIC velocity contribution,0) \\[2mm]
\hspace*{1cm} else if $\nabla\tau_{p}$>0 \\[2mm]
\hspace*{1.7cm}    PIC velocity contribution = max($\delta u_{p}$,$\alpha$*Slip Velocity) \\[1.5mm]
\hspace*{1.7cm}    PIC velocity contribution = min(PIC velocity contribution,0) \\[1.5mm]
\hspace*{1cm} endif

In the logic above, $\alpha$ is a user input applied to solids slip velocity to account for relative dynamics of the neighboring parcels. $\alpha$ is the final model parameter considered for calibration in this study.

For additional information on MFiX-PIC on the implementation of PIC methodology in MFiX-PIC, the reader is referred to~\cite{PICTheoryClarke2020}.

\hypertarget{methods}{\section{Methodology \& 
Software Framework Employed}
\label{methods}}

Several advanced UQ methods and UQ software toolkits were used in the MFiX-PIC calibration analysis presented in this report.  As such, this section provides a brief overview of these methods and toolkits.  Note that the intent of this report does not include teaching the reader the theoretical underpinnings of statistical analysis as it relates to calibration, nor how to use associate software. Therefore, in each subdivision of this section, additional references are provided to direct the user to more comprehensive guidance, if required.

\subsection{Simulation Campaigns and Surrogate Model Construction}

Calibrating input parameters for computational simulation first requires a user to define \emph{quantities of interest/response variables}.  These are measurable values that universally define the accuracy of a simulation.  There may be many input parameters that affect these quantities of interest, and the effect of changing those parameters may be interrelated.  For example, calibrating five input parameters for a single response variable might require thousands of evaluations to find an optimal set of parameters that yield the smallest residual between a simulated and experimental quantity of interest.  To avoid running these thousands of simulations, it is common to construct a surrogate model (a.k.a. a response surface or meta-model) and use it to predict simulation outcomes instead.

Surrogate models are numerous and vary in form and function. In this study, a data-fitted surrogate model, which characterizes the relationship between a response variable and input parameters through sampling simulations that span user prescribed ranges of input parameters was created.  In this work, the language \emph{simulation campaign} describes carefully designed samples 
of simulations, chosen to create a numerical relationship between input parameters and a response variable. In this approach, the simulation code (e.g., MFiX-PIC) is treated as a black box and executed for each sampling simulation as part of a larger predetermined simulation campaign.

Intuitively it seems the number of sampling simulations in a simulation campaign must play a critical role in constructing a reliable data-fitted surrogate model.  In fact, simulation campaigns are designed using a mathematically defined space-filling property to assure enough sampling points within the range of each input parameter are represented.  One common sampling method for computational experiments is Latin Hypercube (LH) sampling \citep{viana2013}.  In this study, a particular Optimal Latin Hypercube (OLH) sampling method is employed whereby a distance metric effectively distributes input parameters to fully span user-defined ranges while ensuring samples are located far from each other.

The workflow outlined below was followed to design the simulation campaign and to construct the data-fitted surrogate models:

\begin{enumerate}
\item Identify the model input parameters to be varied systematically as part of the sampling simulations, and the quantities of interest to be extracted from the results. To bring all stakeholders together and to minimize future disagreements, a survey \citep{Gel2018}  was employed to capture detailed information from the researchers,   
subject matter experts, and other stakeholders involved. After several iterations, the survey provided a clear picture of critical issues, such as how many input parameters would be explored and what the lower and upper bounds of these parameters would be within the simulation campaign.
\item Design the simulation campaign employing OLH sampling principles. In this case, six model input parameters were identified within certain ranges. Although the simulation campaign was carried out for three quantities of interest, for deterministic calibration, only the second quantity of interest (i.e., location of filling shock) was considered.
 Using 20 samples per input parameter, an initial simulation campaign of 120 samples was designed.
\item Launch and monitor the simulation campaign on the targeted HPC system. 
\item Post-process the results from simulations to construct a tabular dataset where each row shows the six model parameter settings and simulation results for each of the quantities of interest corresponding to that sampling simulation. The post-processed dataset is saved as an ASCII file, which consists of the design of experiments for the model parameters and the corresponding quantities of interests from the simulation campaign results. 
\item Import the tabulated dataset into the UQ toolkit software employed, and test different surrogate model options to determine the best data-fitted surrogate model for the given dataset using various statistical metrics.   
For example, cross-validation error assessment was employed to assess the quality of the data-fitted surrogate model.
\end{enumerate}

Once a best data-fitted surrogate model was identified, this same surrogate model was used throughout the subsequent calibration process in lieu of further MFiX-PIC simulations.  The construction of the data-fitted surrogate model was the most time-costly part of this calibration effort.  For a detailed discussion related to surrogate model construction, including error minimization, the reader is referred to earlier studies in \cite{Gel2013a}, \cite{Gel2013b}, and \cite{Gel2016}.

\subsection{Sensitivity Analysis}

Sensitivity analysis is one UQ technique employed to address the important question: ``which input parameters have the most influence on a quantity of interest?"  For calibration purposes, sensitivity analysis plays a key role, particularly when the number of input parameters exceeds three. The technique quantitatively determines the most influential parameters for each quantity of interest, and can be used to focus the attention of experimentalists, particularly when resources are limited.  In the current study, sensitivity analysis identified two key model input parameters in addition to a design variable, which was not targeted for calibration.  Had experimental resources been limited, this would have immediately refocused the calibration effort and minimized physical testing.  However, the problem of interest in this study (particle settling) has an analytic solution, so no physical experimentation was necessary, and the full sweep of input parameters identified by stakeholders was investigated.  However, for follow-up cases (fluidized bed and circulating fluidized bed), the experimental dataset will be limited, and sensitivity analysis is expected to play an important role in guiding the calibration efforts. Hence, the methodology is introduced here.

The sensitivity analysis results shown later in this report (Figure~\ref{fig:C1_R2_rssoboltsi}) were obtained using the Sobol' Indices-based global sensitivity method, which is the preferred methodology for cases with non-linear response behavior. The data-fitted surrogate model was used to perform function evaluations for computing the quantity of interest when calculating the Sobol' indices. 
The reader is referred to \cite{Sobol2001} and \cite{Iooss2015review} for additional information on the methodology, ~\cite{Gel2013a} and \cite{Gel2013b} for a demonstration with multiphase flow simulations. Additionally, a detailed sensitivity analysis study performed for the problem of interest with Nodeworks software can be found in \cite{Vaidheeswaran2020}.

\subsection{Calibration}

Computational models often incorporate empirical input parameters as well as physically observable input parameters.  For example, in MFiX-PIC, only close packed volume fraction would be considered physically observable; all other input parameters are empirical.
The intent of calibration is to tune input parameters with the aid of observable data (e.g., experiments) so that a computational model reproduces expected physics in simulations.
 
\begin{figure}[h]
    \begin{center}
    \includegraphics[clip, trim=0.0in 0.0in 0.0in 0.0in, height=0.15\textheight]{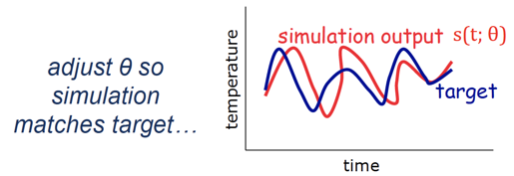}
    \end{center}
    \caption{Illustration of computational model calibration from DAKOTA Training on Calibration \citep{DKTrainCalib}.}
    \label{fig:CalibSketch}
\end{figure}

Figure~\ref{fig:CalibSketch} shows a simple sketch to illustrate the objective of calibration \citep{DKTrainCalib}. For this example, assume the transient temperature behavior in a fluidized bed reactor is being analyzed. The temperature profile in time is represented as the blue line.  This is the target of simulation, most likely observations from sensors or measurements from experiments. Then consider a computational model, $s(t;\theta)$, that aims to capture the temperature behavior in time (red line) through simulation.  The model requires various input parameters, $\theta$, (e.g., heat transfer coefficient).  Recall that most computational models represent a simplification of actual governing physics by employing assumptions, so they will not capture exact physical behavior, hence there is discrepancy between the targeted and simulation results, as illustrated. 

Although some model input parameters might have theoretical foundation, the settings employed for these parameters during simulations are usually considered uncertain. The calibration process aims to minimize the difference between the target and simulation output by adjusting the settings for the $\theta$ parameters.  This is accomplished with the guidance of observations or experimental data representing the target. By reducing the disparity between targeted and simulation results, calibration plays an important role in increasing the credibility of a simulation for a particular application.

At this point, it is important to note the difference between validation and calibration.  Validation is direct comparison of simulation results to experimental results without tuning.  One might use validation to establish a baseline discrepancy between an experiment and a simulation, and use that information to justify the need for model calibration. Both validation and calibration are always performed against a specific set of observable data, which makes the credibility of the experimental data quite critical.  Careful consideration must be given when generalizing the insights gained as a result of calibration studies, particularly when applying previously calibrated input parameters to new simulations. The reader is referred to \cite{Trucano2006} for further information on the difference between validation and calibration.

Typically, there are multiple input parameters (i.e., $\theta_i, i=1,n$) that need to be calibrated concurrently. This situation poses unique challenges especially if experimental data is limited. Consequently, many different calibration approaches are found in literature \citep{DKTrainCalib}.

In general, calibration methods are categorized under two groups: (i) deterministic calibration methods, and (ii) statistical calibration methods. The latter provides a distribution for the calibrated model parameters instead of single values, which is the outcome of deterministic calibration. Another major difference is the ability of statistical calibration to take into account model bias (a.k.a. model form uncertainty) while performing calibration of model input parameters. However, statistical calibration methods usually require the knowledge of complex methods and algorithms such as Markov Chain Monte Carlo (MCMC). On the other hand, deterministic calibration is easier to understand and widely implemented in various software tools that have optimization capability.

\subsection{Deterministic Calibration}

The goal of deterministic calibration is to find values of $\bm{\theta} : \{\theta_1 \dots \theta_m\}$ that will minimize residual error between a group of simulations and their equivalent experimental counterparts.  Eqn. \ref{eqn:minimizeRes} acts as the objective function for the optimization problem \citep{DKTrainCalib}. It represents the sum of squares of the residual errors introduced by employing this set of $\bm{\theta}$ in $n$ simulations.

\begin{equation}
\underset{\bm{\theta} \in \mathbb{R} }{\text{minimize}} f(\bm{\theta}) =  \sum_{i=1}^n {\left[ S_i (\bm{\theta}) - y_i \right]}^2 = \sum_{i=1}^n {\left[ R_i (\bm{\theta}) \right]}^2
\label{eqn:minimizeRes}
\end{equation}
\hspace*{2.15cm} where $\bm{\theta} = \{\theta_1 \dots \theta_m\}$ are the modeling parameters being calibrated \\
\hspace*{3.15cm} $y_i$ is the $i^{th}$ experiment data observed out of $n$ experiments\\
\hspace*{3.15cm} $S_i(\bm{\theta})$ is the simulation result for $i^{th}$ experiment data as function of $\theta_1 \dots \theta_m$ \\
\hspace*{3.15cm} $R_i(\bm{\theta})$ is the $i^{th}$  residual (simulation - experiment)  \\
\hspace*{3.15cm} $\mathbb{R}$ is the real numbers for the modeling parameters  \\

Depending on the nature of the problem there are various local and global optimization techniques that could be employed to solve the residual minimization problem shown in Eqn. \ref{eqn:minimizeRes}. In this light, an important distinction between statistical calibration and deterministic calibration is that the outcome from statistical calibration is an estimated distribution of the $\bm{\theta} $ parameters individually, whereas deterministic calibration provides a single scalar value for each of the model parameters being calibrated.

\subsubsection{Workflow for Deterministic Calibration}

\begin{figure}[H]
    \begin{center}
    \includegraphics[clip, trim=0.0in 0.0in 0.0in 0.0in, height=0.4\textheight]{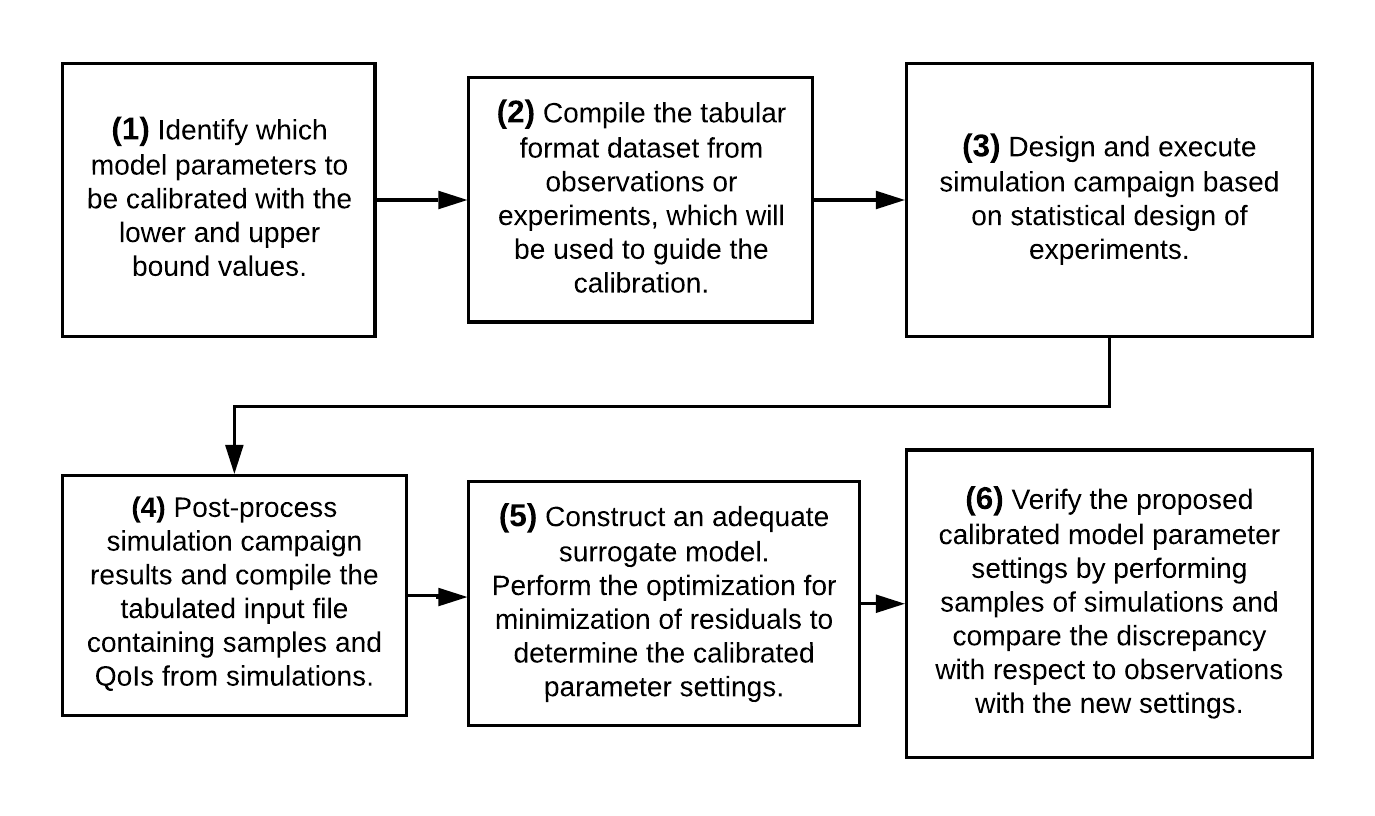}
    \end{center}
    \caption{Illustration of the deterministic calibration workflow performed in this study.}
    \label{fig:CalibWorkflow}
\end{figure}

The workflow outlined below was followed to perform deterministic calibration in this study.  For a visual perspective, the same workflow is illustrated in Figure~\ref{fig:CalibWorkflow}:

\begin{enumerate}
\item Identify the model parameters to be calibrated, and determine the lower and upper bounds for each of these parameters to be used during calibration.
\item Prepare an experimental dataset or observations to be used to guide the calibration process as an ASCII input file.
\item Plan a simulation campaign with the aid of statistical design of experiments principles that will enable the construction of a data-fitted surrogate model.  The surrogate model should adequately characterize the relationship between model parameters considered as input and the response variables (a.k.a. quantities of interest or output). This step is crucial when the simulations are expensive or time consuming to perform as the optimization process requires thousands of function evaluations to be performed at a low computational cost.
\item Post-process the simulation campaign results and compile an ASCII file as a tabulated dataset consisting of the design of experiments for the model parameters and the corresponding quantities of interest from the simulation campaign results. For calibration, a separate dataset containing the experimental observations is necessary. This should be prepared in ASCII format for importing into UQ software. For the current application, an analytical solution was available and used in lieu of experimental observation data. Twenty-one samples were created by varying the control parameter ($x_1$: Initial solid concentration).
\item Utilize UQ toolkit (\href{https://computing.llnl.gov/projects/psuade/software}{\tt PSUADE}, \href{https://mfix.netl.doe.gov/nodeworks}{\tt Nodeworks}) to import the datasets and perform the optimization required to minimize the residuals in Eqn. \ref{eqn:minimizeRes}. The minimization procedure may necessitate multiple attempts which will generate several sets of values for $\theta_i, i=1,n$. Each attempt will yield a minimum residual for all experimental samples. A parallel coordinates plot that incorporates all of the proposed settings of $\theta_i, i=1,n$ is utilized to identify the most frequently encountered values. Note that the surrogate model constructed is used to perform the evaluations required for $S_i (\bm{\theta})$ in Eqn. \ref{eqn:minimizeRes} in lieu of actual MFiX-PIC simulations for each instance. Hence, the credibility of the surrogate model needs to be carefully assessed prior to the optimization step with measures such as adjusted $R^2$ or cross-validation error assessment. Doing so ensures the error introduced by the surrogate is minimized.
\item Verify the proposed calibrated model parameter settings by re-running a select group of simulations within the existing simulation campaign or by constructing a new simulation campaign for unseen samples. In both cases, any error needs to be assessed against an experimental solution to determine whether the calibrated settings truly improve the credibility of the simulation model for the targeted application.
\end{enumerate}

\subsection{Bayesian (Statistical) Calibration}

Bayesian calibration is different than the deterministic calibration methodology presented in the previous section. Bayesian calibration employs Bayes Theorem, which simply relates prior information with associated uncertainty to future information based on the likelihood of observed outputs from the model~\citep{Muehleisen2016}. In the deterministic calibration method, the objective is to find a set of values for the uncertain model parameters that minimize the residual error difference between observed data from experiments and model computed quantities of interest. However, in Bayesian calibration, the objective is to determine the most likely uncertainties for input parameters that yield the quantity of interest, with some uncertainty, in which the observed data is most likely to reside ~\citep{Muehleisen2016}.  It is an iterative process of updating distributions with targeted uncertain parameters in a way that is consistent with observed data.

The Bayesian calibration methodology was first proposed by~\cite{Kennedy_OHagan2001} as a statistical framework where a ``hierarchical model linking noisy field measurements from the physical system to the potentially biased output of a computer model run with the `true' (but unknown) value of any calibration parameters not controlled in the field''~\citep{Gramacy2015}. The backbone of the framework is a pair of coupled Gaussian process (GP) priors for (a) simulator output and (b) bias. For the latter GP model, the systemic differences between the model predictions and observations, which are commonly referred to as ``model error,'' ``model form uncertainty,'' or ``structural error.'' are taken into account~\citep{Maupin2020}. The hierarchical nature of the model, paired with Bayesian posterior inference, allows both data sources (simulated and field) to contribute to joint estimation of all unknowns.

One of the major advantages of Bayesian calibration over deterministic calibration is  ``the ability to retrieve a full description of the uncertainties about the parameters and consequently about the simulator outputs.'' \citep{Guillas2014}.  Moreover, the ability to express ``uncertain'' scientific beliefs regarding the model parameters in terms of ``priors'' within Bayesian framework enables a natural integration of scientific knowledge and evidence given by measurements. \citep{Guillas2014}. On the other hand, a major disadvantage of Bayesian calibration is the inherently iterative nature of the calibration process, which may require a substantial number of iterations to converge to the most likely posterior distribution. Such long iteration counts are typically encountered if the prior distributions are poorly chosen, and thus, require significant updating during the calibration process~\citep{Muehleisen2016}. Hence, careful selection of the priors is quite important and there are numerous studies in the literature specifically investigating the effect of different priors for Bayesian calibration. For example, in the study of ~\cite{Ling2014}, Bayesian calibration with several different prior formulations of the model discrepancy function (model bias) ranging from constant to Gaussian random process with non-stationary covariance function were investigated. Another disadvantage of Bayesian approach is ``identifiability'', i.e., the ability to distinguish between uncertain model input parameters (which is the primary motivation for calibration) and systematic inadequacies (model bias), which is often quite challenging ~\citep{arendt2010}. For additional information on these issues, the reader is referred to ~\cite{Maupin2020},\cite{wang2021}, and \cite{Ling2014}.

As illustrated in \cite{Guillas2014}, the Bayesian calibration process allows one to : (i) evaluate a small systematic bias of a CFD model; (ii) narrow down the set of parameter values that provides the best match between CFD model outputs and the observations (i.e., the analytical solution in this case); (iii) construct a cheap-to-evaluate statistical surrogate model (also called emulator) of the CFD model; and (iv) use the surrogate model to quantify the uncertainty of the quantity of interest (i.e., location of the settling shock) resulting from both uncertainties in the MFiX-PIC model parameters and the numerical code itself (due to various assumptions and simplifications), as well as measurement errors. This propagation of uncertainties would be a very computationally demanding task without the use of a surrogate model.

\begin{figure}[H]
    \begin{center}
    \includegraphics[clip, trim=0.0in 0.0in 0.0in 0.0in, height=0.15\textheight]{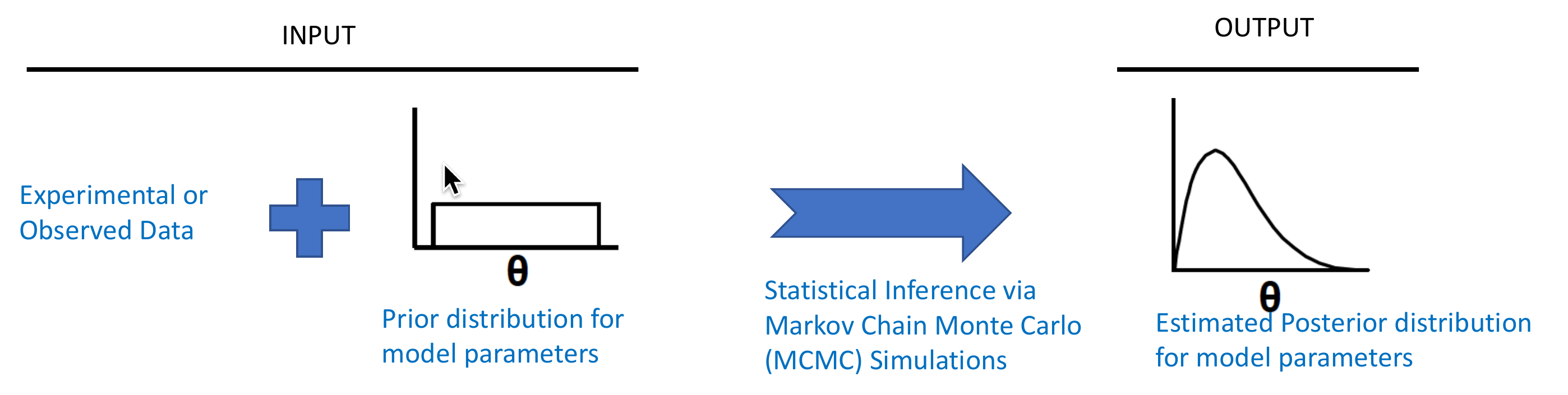}    
    \end{center}
    \caption{Schematic illustration of the Bayesian calibration \citep{DKTrainCalib}.}
    \label{fig:Sketch_BayesianCalib_DK}
\end{figure}

Figure~\ref{fig:Sketch_BayesianCalib_DK} shows a high-level illustration of the Bayesian calibration framework, which starts with a prior distribution of the model parameters based on current beliefs. Hence, the assumptions for prior distribution makes a difference. Then observations from the experiments are employed to guide the calibration process which simply employs Bayes Theorem to estimate the posterior distribution of the model parameters. The Markov Chain Monte Carlo (MCMC) based approach is employed to perform Bayes's rule. 

The reader is referred to~\citep{VianaSurvey2021} for a comprehensive literature review of Bayesian calibration methods and their applications.

In this study, the original Bayesian calibration framework from \cite{Kennedy_OHagan2001} was followed. It is based on representing model bias and quantities of interest from the computer model as Gaussian processes, to investigate how the sophisticated tuning of the five MFiX-PIC model parameters can improve the prediction accuracy of the location of the settling shock. A major deviation from the original Bayesian calibration framework was the availability of the observations to guide the calibration. Typically, experimental data with some uncertainty is utilized whereas an analytical solution was employed, and artificial uncertainty (less than 1\%) was introduced to characterize the effect of experimental uncertainty.

Following the notation in \cite{VianaSurvey2021}, if we assume the simulator (MFiX-PIC) sufficiently represents the physical system that the observations are taken from, then the relationship between the observations and simulator can be expressed in the following equation:
\begin{equation}
y(\mathbf{x}_i, t) = \eta(\mathbf{x}_i, t; \theta) + \delta(\mathbf{x}_i, t) + \epsilon_i
\end{equation}
\hspace*{9mm} where $\mathbf{x}_i$ are the input settings such as geometry and directly measurable variables; \\
\hspace*{19.45mm} (e.g., initial solids concentration, $\epsilon_{s0}$ in this case ) \\
\hspace*{17.25mm} $i$ shows the $i^{th}$ observation out of $n$ observations (experiments with uncertainty); \\
\hspace*{17.25mm} $t$ denotes time for temporally transient problems; \\
\hspace*{17.25mm} $y$ are the output observations (in this case analytical solutions to guide calibration) ;  \\
\hspace*{17.25mm} $\eta(\mathbf{x}_i, t; \theta) $ is the simulator (in this case MFiX-PIC simulations);  \\
\hspace*{17.25mm} $\theta$ is the set of calibration parameters that cannot be directly measured \\
\hspace*{19.45mm} (e.g., $\theta_1$:P\_0  Pressure linear scale factor, $\theta_5$:VelfacCoeff ); \\
\hspace*{17.25mm} $\delta(\mathbf{x}_i, t) $ is a stochastic term that accounts for discrepancy between simulator \\
\hspace*{19.45mm} and reality (a.k.a. model form uncertainty or model bias) \\
\hspace*{17.25mm} $\epsilon_i$ represents the variability in observation.

Bayesian calibration starts with a prior distribution of the uncertain model parameters, which reflects our beliefs about the parameters. Then Bayes Theorem is employed to update
initial beliefs in the model parameters with the help of observations. Hence, the posterior distribution of the model parameters can be calculated as the solution of Eqn.~\ref{posterior} for a given set of observations and simulator results, i.e., $\mathbf{z} = [y^T , \eta^T ]$~\citep{VianaSurvey2021}:
\begin{equation}
p( \theta, \mu, \lambda, \beta | \mathbf{z}) = \frac{L(\mathbf{z} |  \theta, \mu, \lambda, \boldsymbol{\beta}, \sum) \, p_0 (\theta) \, p_0 (\mu, \lambda, \boldsymbol{\beta})}{\int L(\mathbf{z} |  \theta, \mu, \lambda, \boldsymbol{\beta}, \sum) \, p_0 (\theta) \, p_0 (\mu, \lambda, \boldsymbol{\beta}) \, d\theta d \mu d \lambda d \beta}
\label{posterior}
\end{equation}
\hspace*{10mm} where \\
\begin{equation*}
L(\mathbf{z} |  \theta, \mu, \lambda, \boldsymbol{\beta}, \sum) = {| \sum | }^{-1/2} exp \left[ -\frac{1}{2} ( \mathbf{z} - \mu )^{T} \, {\sum}^{-1} ( \mathbf{z} - \mu ) \right]
\end{equation*}
\begin{equation*}
\sum = {\sum}_{\eta} + 
\begin{bmatrix}
\sum_y & 0 \\
0 & 0 
\end{bmatrix}
\end{equation*}
\hspace*{19.25mm} $\mu, \lambda, \boldsymbol{\beta}$ are the hyperparameters of the Gaussian process models; \\
\hspace*{19.25mm} $L(.)$ is the likelihood of observed data given the set of hyperparameters;\\
\hspace*{19.25mm} $p_0$ are prior distributions for the hyperparameters; \\
\hspace*{19.25mm} $\sum$ is the covariance matrix (defined in terms of parameters $\lambda$, and $\beta$)

The above equation is known to be intractable and numerical integration is usually carried out using a MCMC method  \citep{VianaSurvey2021}, which is one of the reason for Bayesian calibration to be computationally expensive. Solving Eqn.~\ref{posterior}  returns the joint posterior distribution of parameters $\theta$, $\mu$, $\lambda$, and $\beta$. Therefore, besides estimate uncertainties in the calibration parameters $\theta$, one can also perform predictions at points that were not previously observed (simulations or experiments) using the Gaussian Process models constructed.
For additional details of the above formulation, the reader is referred to the survey paper by ~\cite{VianaSurvey2021}.

\subsubsection{Workflow for Bayesian Calibration}

Following the outline in ~\cite{Muehleisen2016}, the general workflow for Bayesian calibration can be described under three major steps:
\begin{enumerate}
\item Define prior distributions based on the beliefs about uncertain model parameters (e.g., characterize the uncertainty for $\theta_1$ to $\theta_5$ with assumed PDFs, which for this case was uniform distribution within a prescribed lower and upper bounds).
\item Collect experimental observations based on the design variables (e.g., for this case $x_1$: initial concentration was varied to compute the location of settling shock from the analytical solution, which was used in lieu of the experiments).
\item Calibrate (assumed) prior parameter PDFs based on the observed data by iteratively using Bayes' Theorem until iterations converge to an acceptable level \citep{Kennedy_OHagan2001}, which yields the estimated posterior distribution of the model parameters considered for calibration (i.e., $\theta_1$ to $\theta_5$).
\end{enumerate}

Due to the computational complexity and intensive resource requirements of the above workflow, an open-source UQ toolkit which can partially automate the above steps has been employed in this study. A high level overview of the UQ toolkits available with Bayesian Calibration features are presented in the next section.

\subsection{Software Toolbox Employed}
There are several open-source UQ software that can be employed for the calibration study presented in this report such as PSUADE~\citep{PSUADE,PSUADEweb}, DAKOTA~\citep{DAKOTA2008,DAKOTA}, etc. In the previous report~\citep{Gel2021TRS}, for the deterministic calibration study, both PSUADE and DAKOTA were used in addition to the NETL-developed tool Nodeworks~\citep{weber2020}. 
However, among these three software, Bayesian calibration capability was only available within PSUADE and DAKOTA. The latter required more advanced setups and also several issues were encountered when constructing a surrogate model concurrently during Bayesian calibration. These issues motivated the authors to use only PSUADE for the presented Bayesian calibration study with comparison against the PSUADE-based deterministic calibration results. A brief description of the PSUADE software tool is presented in this section.

\subsubsection{PSUADE:} \href{https://computing.llnl.gov/projects/psuade/software}{PSUADE} is an open-source UQ software toolkit developed at the Lawrence Livermore National Laboratory ~\citep{PSUADE} and released under LGPL license since 2007. The name of the software,  PSUADE, comes from the acronym for \underline{P}roblem \underline{S}olving Environment for \underline{U}ncertainty \underline{A}nalysis and \underline{D}esign \underline{E}xploration. The program supports a variety of non-intrusive uncertainty quantification analysis methods where the simulation application can be treated as ``black-box" code.  Subsequently, many UQ analysis tasks can be performed by sampling the black-box directly or through a data-fitted surrogate model constructed from the computational model. The software offers a diverse range of sampling methods to enable users to perform simulation campaigns with the objective of constructing an adequate data-fitted surrogate model (a.k.a. response surface model, meta-model). The user can perform both basic uncertainty analysis such as forward propagation of uncertainties and more complex analysis such as mixed aleatory-epistemic uncertainty analysis. PSUADE has a built-in statistical calibration capability (i.e., Bayesian calibration with MCMC). 
However, deterministic calibration required user-defined supporting code to incorporate residual evaluations.  PSUADE is written in C++ and operates primarily as a command-line based software, which may require some learning curve. Additional details on the capabilities of PSUADE can be found at the website of the software~\citep{PSUADEweb}.

\hypertarget{demonstration}{\section{Calibration Demonstration Cases}
\label{demonstration}}


\hypertarget{Overview of Demonstration Cases Considered}{\subsection{Overview of Demonstration Cases Considered}\label{DemoCase}}

This report is a continuation of the previous efforts in documenting a series of calibration studies that aimed to investigate three industrial applications, which span a wide range of flow regimes.  In particular, the cases of interest were particle settling, a fluidized bed, and a circulating fluidized bed.  The first report \citep{Gel2021TRS} concentrated on the particle settling problem, in the context of deterministic calibration. The current report aims to complement the previous effort by demonstrating application of Bayesian calibration (a.k.a. statistical calibration) to the same problem and comparing the improvements achieved by both approaches. In addition, a bug was discovered in MFiX-PIC after the publication of the first report \citep{Gel2021TRS}, which was determined to have an effect on the deterministic calibration results presented earlier. The updated results for the deterministic calibration study are also presented and used in the comparison with Bayesian calibration results.

The objective of a deterministic calibration study is to obtain a set of optimal model parameters for a given application problem.  Similarly, the objective for Bayesian calibration is to estimate the likely distribution of the uncertain model parameters inferred from the experimental data available instead of scalar values of the parameters as done in deterministic calibration.
Ultimately, the goal of the authors of this report is to provide the results of the Bayesian calibration proposed MFiX-PIC settings and compare those against the results of deterministic calibration to assess which calibration approach offered the most improvement of the accuracy of MFiX-PIC for this category of flow problem.

\subsection{Gravitational Particle Settling}
\begin{figure}[H]
    \begin{center}
    \includegraphics[clip, trim=0.0in 0.0in 0.0in 0.0in, height=0.2\textheight]{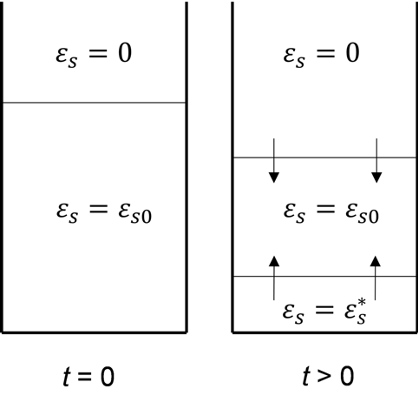}
    \end{center}
    \caption{Schematic of particles settling in a dense medium.}
    \label{fig:C1schematic}
\end{figure}

Particles settling in a dense medium under gravity is suitable for calibration since it has an analytical solution for the QoI (e.g., location of the filling shock). This eliminates the uncertainty associated with measurements from experiments. An added advantage is that the QoI has an algebraic form, whereby additional data may be generated easily. The setup and operating conditions follow the work of ~\citet{MFIXPICVV} as illustrated in Figure~\ref{fig:C1schematic}. The domain is 0.02 m $\times$ 1 m $\times$ 0.02 m in x, y and z directions, respectively. The simulations are run at the sampling locations for a duration of 1 second using a constant time-step size of 5e-4 s.

\begin{figure}[htb]
    \begin{center}
    \includegraphics[clip, trim=0.0in 0.0in 0.0in 0.0in, width=0.3\textheight]{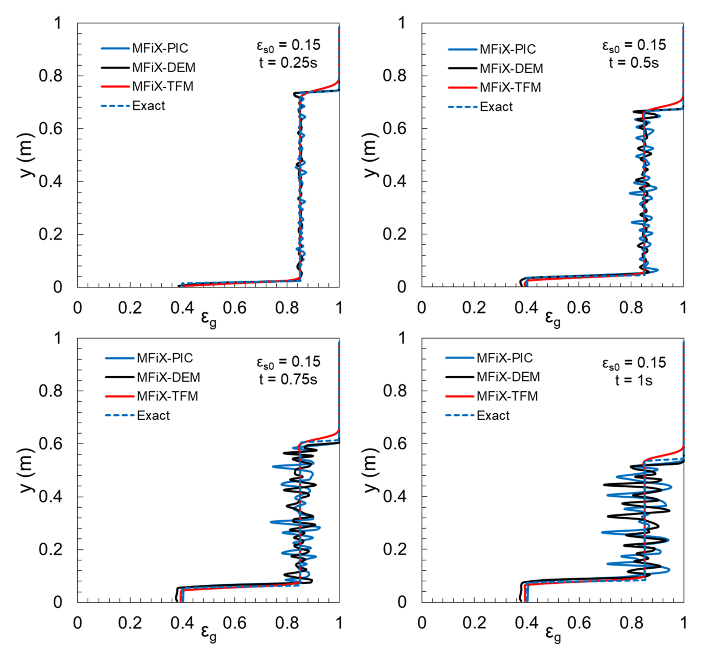}
    \end{center}
    \caption{Comparison of time evolution of shock fronts obtained using uncalibrated MFiX-DEM, MFiX-PIC and MFiX-TFM simulations with the analytical solution.}
    \label{fig:C1MFIXComp}
\end{figure}

Once the simulation begins, two concentration (kinematic) shocks evolve. The first originates from the top of the particle bed and corresponds to settling, while the other originates from the bottom of the vessel and corresponds to a filling shock. The location of the filling shock ($y_2$) is the QoI considered in this study. Its analytical solution is given by:

\begin{equation}
y_2(t) = -t  \left( \frac{\epsilon_s^* \epsilon_g^* u_r^* - \epsilon_{s0} \epsilon_{g0} u_{r0}}{\epsilon_s^* - \epsilon_{s0}} \right)
\label{eqn:analytical}
\end{equation}

where $\epsilon_{s}^{*}$ and $\epsilon_{g}^{*}$ represent volume fractions of solids phase and gas phase at close-packing conditions. $\epsilon_{s0}$ and $\epsilon_{g0}$ represent initial volume fractions. $u_{r0}$ and $u_r^{*}$ are relative velocities calculated using initial and close-packing conditions, respectively. This specific case is comparable to parcels having a relatively slower dynamics for instance, as observed in a standpipe of a circulating fluidized bed or in a moving bed reactor. This case was used in the Verification and Validation Manual ~\citep{MFIXPICVV} for comparing results from MFiX-DEM, MFiX-PIC and MFiX-TFM as shown in Figure~\ref{fig:C1MFIXComp}.

\subsection{Simulation Campaigns and Surrogate Models}

Several simulation campaigns were designed and executed to create data-fitted surrogate models that can adequately characterize the relationship between input parameters and the quantities of interest from MFiX-PIC. A brief description of the simulation campaigns used for the presented calibration study is provided below:
\begin{itemize}
\item {\bf Simulation Campaign \# 1:} Targeted primarily for the deterministic calibration study. For the initial calibration study, a simulation campaign with a sample size of 120 MFiX-PIC  simulations was employed. This sample size corresponds to the double of the commonly accepted guidance of 10 samples per uncertain parameter. The additional samples were aimed  to capture the relationship between the input parameters and QoI better. However, a bug fix in MFiX-PIC was discovered after the initial calibration study results was published~\citep{Gel2021TRS}. The bug fix made the calculation of {\tt ROP\_g} (product of gas density times gas volume fraction) consistent with {\tt EP\_g} (gas volume fraction) so the updates to the two variables were synchronized~\citep{MFIXBugFix21_2Release}. However, the bug fix affected the results from our earlier simulation campaigns used to construct the surrogate models generated in \cite{Gel2021TRS}, which had to be rerun. After re-running the original 120 sample simulation campaign, several revisions in the lower and upper bounds were deemed necessary which forced a new simulation campaign to be constructed. Based on the insight gained from the previous campaigns, a new simulation campaign with the half of the original sample size was constructed to test if the campaign size footprint could be reduced while achieving similar outcomes in terms of surrogate model quality. Hence, the total sample size for this campaign was 60 samples instead of 120 samples, which turned out to enable constructing an adequate quality data-fitted surrogate model that will be sufficient for the deterministic calibration study of the particular problem configuration. As mentioned in earlier reports, the sample size is one of the important parameters in the quality of the constructed data-fitted surrogate model. However, a separate study is planned similar to the one presented in the original report~\citep{Gel2021TRS} to assess the sensitivity of the deterministic calibration results on the surrogate model constructed from 60 samples versus 120 samples.
\item {\bf Simulation Campaign \# 2:} Targeted for the Bayesian calibration study with an initial 120 sampling simulations, which was later augmented with 64 additional samples.
\end{itemize}
The surrogate models constructed from the simulation campaign results were then used in lieu of actual MFiX-PIC simulations to provide cheaper evaluations of the quantity of interest needed during the calibration study. The quality of the constructed surrogate model plays an important role as its evaluations for the quantity of interest must be as close as possible to an MFiX-PIC simulation result.

\subsubsection {Simulation Campaign \# 1}
\paragraph{Design of Sampling Simulations:} 
A detailed description of the process for designing the sampling simulation campaigns was provided in the prior deterministic calibration study~\citep{Gel2021TRS}. Following the same approach, OLH sampling method has been used to generate a simulation campaign for six MFiX-PIC input parameters.
The first five were modeling parameters specific to MFiX-PIC, accessible to the user through keywords.  These included: $\theta_1:$ Pressure linear scale factor $(P_0)$; $\theta_2:$ Volume fraction exponential scale factor $(\beta)$; $\theta_3: $Statistical Weight $(W_p)$; $\theta_4:$ Void fraction at maximal close packing $(\epsilon_g^*)$; and $\theta_5: $ Solids slip velocity scale factor $(\alpha)$.  The sixth parameter was initial solids concentration, a general input parameter used to specify an initial condition in MFiX, regardless of model.
In the remainder of this report, abbreviated versions of the input parameter names might be used due to font issues in plotting software. Table~\ref{tb:InputParams_OLHn60} offers these abbreviations along with lower and upper bound values assumed for each model parameter in the simulation campaign. For example, anywhere t1:P\_0 or t1 or Theta1 appears in this report, it is equivalent to $\theta_1$:Pressure linear scale factor $(P_0)$.

\begin{table}[htb]
\begin{center}
\begin{tabular}{|l|l|l|r|r}
\cline{1-4}
\cellcolor[HTML]{00D2CB}{\color[HTML]{333333} \textbf{Symbol}} & \cellcolor[HTML]{00D2CB}{\color[HTML]{333333} \textbf{Description}} &  \cellcolor[HTML]{00D2CB}{\color[HTML]{333333} \textbf{Min.}} &  \cellcolor[HTML]{00D2CB}{\color[HTML]{333333} \textbf{Max.}}  \\ 
\cline{1-4}
$\theta_1$ or t1:P\_0  & Pressure linear scale factor, $(P_0)$       & 0.488  & 19.99 \\ 
\cline{1-4}
$\theta_2$ or t2:beta & Volume fraction exponential scale factor, $(\beta)$   & 2.0 & 5.0 \\
\cline{1-4}
$\theta_3$ or t3:StatWeight & Statistical Weight, ($W_p$)   & 3.0  & 15.0 \\  
\cline{1-4}
$\theta_4$ or t4:ep\_g*  & Void fraction at maximal close packing, $(\epsilon_g^*)$  & 0.38  & 0.43 \\ 
\cline{1-4}
$\theta_5 $ or t5:VelfacCoeff  & Solids slip velocity scale factor, $(\alpha)$   & 0.5 & 0.9 \\ 
\cline{1-4}
$x_1$          & Initial solids concentration, ($\epsilon_{s0}$)  & 0.05  &  0.25 \\ 
\cline{1-4}
\end{tabular}
 \caption{List of input parameter abbreviations, descriptions, lower and upper bounds values considered in simulation campaign \# 1 with 60 samples.}
    \label{tb:InputParams_OLHn60}
\end{center}
\end{table}

\vspace*{-4mm}
The initial set of quantities of interest (a.k.a. response variables) retrieved from the simulation campaign results included, $y_1$:Location of Settling Shock; $y_2$:Location of Filling Shock; and $y_3$:Void fraction in the first cell nearest to the bottom of the experimental vessel. The scope of the work presented herein is to analyze the performance of MFiX-PIC in regions having dense concentration of particles. Because of this, only $y_2$ is used in this analysis to be consistent with the deterministic calibration study. 
Hence, only $y_2$:Location of Filling Shock was considered the key quantity of interest for calibration purposes. 
Note that in the remainder of this report, abbreviated versions of this quantity of interest name might be used due to font issues in plotting software such as ``y2:LocSettling'' which corresponds to $y_2:$Location of Filling Front or Shock.

\begin{figure}[htp]
    \begin{center}
    \includegraphics[clip, trim=0.0in 0.0in 0.0in 0.0in, height=0.4\textheight]{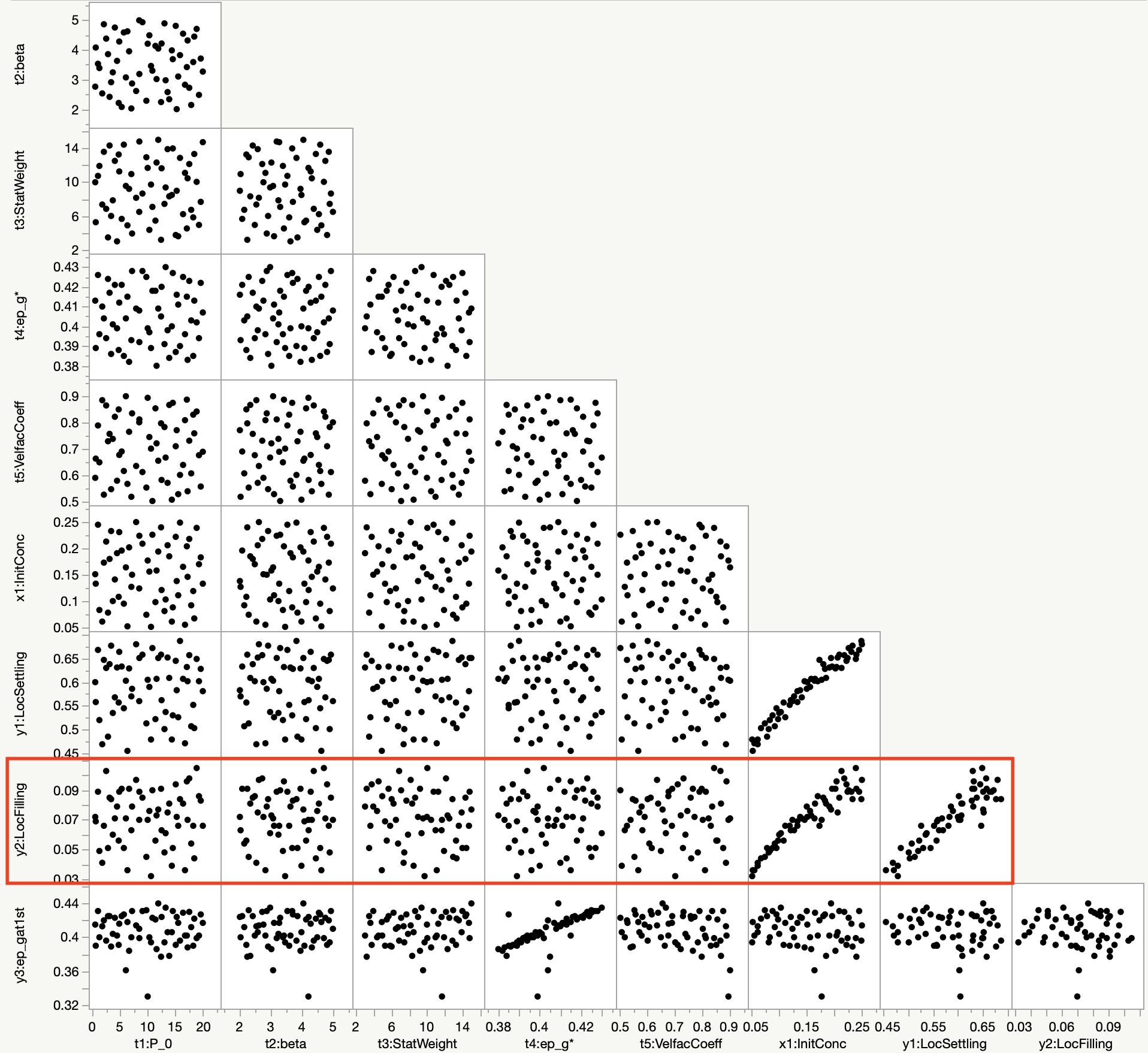}
    \end{center}
    \caption{Scatter matrix plot of all input parameters and all QoI employed in the simulation campaign \# 1 using OLH sampling (60 samples). Note that only the row highlighted in red rectangle is used in the study for the quantity of interest (i.e., y2:LocSettling). }
    \label{fig:SIM7_n60_i6_o3_scattermatrix}
\end{figure}

Figure~\ref{fig:SIM7_n60_i6_o3_scattermatrix} shows a scatter matrix plot of all input parameters and all of the output variables retrieved (a.k.a. quantities of interest) from the simulation campaign. Although the scatter matrix plot shows three quantities of interest retrieved (i.e., $y_1$:Location of Settling, $y_2$:Location of Filling Shock, $y_3$:Void fraction in the first cell nearest to the bottom of the experimental vessel) , the primary focus of the current calibration study was only $y_2$:Location of Filling Shock. The associated row in the scatter matrix plot is highlighted with a red rectangle box. This type of image can be used to make a quick visual assessment of obvious correlations. For example, Figure~\ref{fig:SIM7_n60_i6_o3_scattermatrix} indicates that there is a strong linear correlation between Initial Concentration ($x1$, on horizontal-axis) and Location of Settling Shock ($y1$, on vertical-axis);  this evaluation is based on examining the block representing ($x1$ v. $y1$) as an independent graph and noting a generally linear correspondence between the variables.

\subsubsection{Surrogate Model Construction:} 
The next step in the workflow is to construct a surrogate model from the simulation campaign results, which will be used in lieu of the MFiX-PIC simulations for cheaper evaluations of the quantity of interest when performing UQ analysis or sensitivity analysis.

For this purpose, several surrogate model types were evaluated during the construction process such as Multivariate adaptive regression spline (MARS), linear regression and the Gaussian process model (GPM).  The objective is to find the best data-fitted surrogate model that characterizes the relationship between the given input parameters and quantity of interest from the simulation campaigns.

\begin{figure}[htp]
    \begin{center}
    \includegraphics[clip, trim=0.0in 0.125in 0.0in 0.1in, height=0.33\textheight]{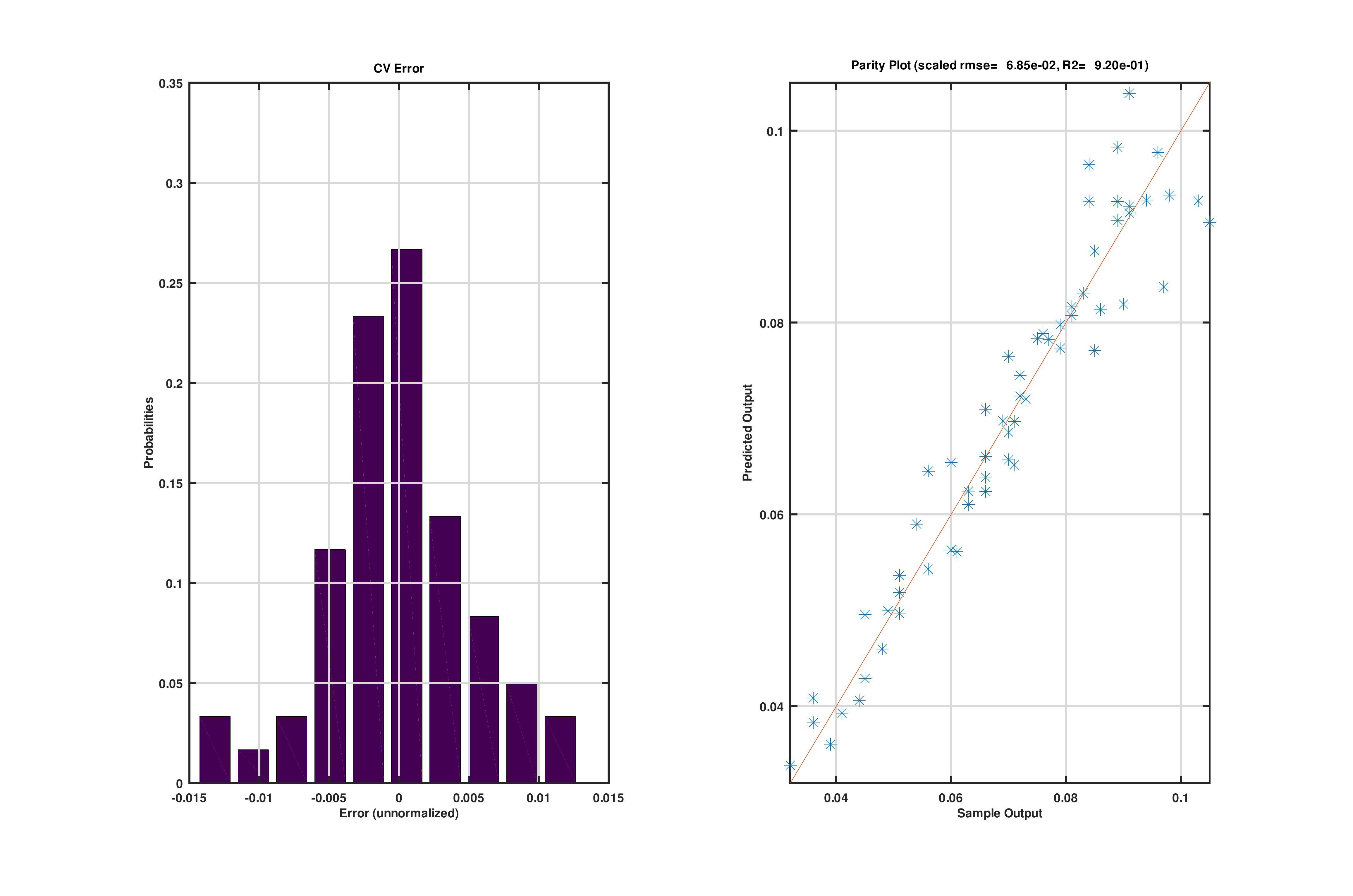} 
    \end{center}
    \caption{Assessment of the quality of the surrogate model (based on 60 samples) through cross-validation errors.}
    \label{fig:C1_n60_RSFA_cv_err}
\end{figure}

In the end, a Gaussian process model, which was a Tong implementation of GPM in PSUADE~\citep{PSUADE,PSUADEweb} with option 10 for RSM, appeared to provide one of the best fits based on the unscaled RMSE, which was calculated by PSUADE to be 5.188e-03  
for the data-fitted surrogate model. 
Figure~\ref{fig:C1_n60_RSFA_cv_err} shows cross-validation error assessment results created with PSUADE. The parity plot on the right compares
actual MFiX-PIC simulation results (horizontal axis with “Sample Output” label) with the surrogate model’s predictions
(vertical axis with “Predicted Output” label). Ideally, sample points (shown as blue asterisk symbols) should fall along
a 45◦ line through the parity plot. Any deviation from the diagonal line reveals discrepancy between the simulation
and data-fitted surrogate model, which implies an additional level of uncertainty being introduced when the surrogate
model is used in lieu of the corresponding MFiX-PIC simulations. To better illustrate the error between simulation and
data-fitted surrogate model, the histogram on the left reveals how the deviation from the diagonal is distributed. Ideal
distribution of errors is expected to be centered around zero, have a narrow spread and tails without any skew.
The input file for the PSUADE data fitted surrogate model is provided in Appendix~\ref{appendix:PS_RSM}. 

\subsubsection {Simulation Campaign \# 2}
\paragraph{Design of Sampling Simulations:} 

The steps for simulation campaign \# 2 are similar to the first campaign, except the campaign was initially designed with 120 sampling simulations. However, the lack of adequate samples at the boundaries of the parameter space resulted in several posterior distributions of calibrated model parameters showing peaks at both lower and upper bounds. This is possibly a sign that the surrogate model has large errors near the edge of the parameter input space. To address this issue, the OLH sampling-based 120 samples were augmented with factorial design-based 64 samples to capture the QoI behavior more accurately within the parameter space boundaries. On a separate note, the 60 sample-based simulation campaign used in campaign \#1 was not available at the time Bayesian calibration studies launched. Hence, the 120 sample-based campaign was utilized. However, a separate study is planned to check if the same issue mentioned above will be observed (i.e., posterior distributions peaking at the boundaries of the parameter space). A new set of samples were created for the 60 sample campaign, which may not suffer from the same issue observed in terms of availability of adequate number of samples close to the bounds of the parameter space.

Table~\ref{tb:InputParams_OLHn120} shows the five model parameters and the single physical variable lower and upper bounds. The bounds may show slight differences from those presented in Table~\ref{tb:InputParams_OLHn60} due to adjustments done after the initial samples are generated in a non-dimensional way (i.e., ranges set between 0 to 1) and then mapped to the actual parameter lower and upper bounds. The Latin Hypercube sampling does not ensure a sample will exist on the desired bound value, therefore, sometimes it is better to increase the bounds with a slight offset value to ensure at least one sample with the targeted bound value exists.

\begin{table}[htp]
\begin{center}
\begin{tabular}{|l|l|l|r|r}
\cline{1-4}
\cellcolor[HTML]{00D2CB}{\color[HTML]{333333} \textbf{Symbol}} & \cellcolor[HTML]{00D2CB}{\color[HTML]{333333} \textbf{Description}} &  \cellcolor[HTML]{00D2CB}{\color[HTML]{333333} \textbf{Min.}} &  \cellcolor[HTML]{00D2CB}{\color[HTML]{333333} \textbf{Max.}}  \\ 
\cline{1-4}
$\theta_1$ or t1:P\_0  & Pressure linear scale factor, $(P_0)$       & 0.488  & 20.0 \\ 
\cline{1-4}
$\theta_2$ or t2:beta & Volume fraction exponential scale factor, $(\beta)$   & 2.0 & 5.0 \\
\cline{1-4}
$\theta_3$ or t3:StatWeight & Statistical Weight, ($W_p$)   & 2.95  & 15.0\\  
\cline{1-4}
$\theta_4$ or t4:ep\_g*  & Void fraction at maximal close packing, $(\epsilon_g^*)$  & 0.38  & 0.4299 \\ 
\cline{1-4}
$\theta_5 $ or t5:VelfacCoeff  & Solids slip velocity scale factor, $(\alpha)$   & 0.5 & 0.899 \\ 
\cline{1-4}
$x_1$          & Initial solids concentration, ($\epsilon_{s0}$)  & 0.05  &  0.2498 \\ 
\cline{1-4}
\end{tabular}
 \caption{List of input parameter abbreviations, descriptions, lower and upper bounds values considered in simulation campaign \# 2 with 120+64 samples.}
    \label{tb:InputParams_OLHn120}
\end{center}
\end{table}

\begin{figure}[htp]
    \begin{center}
    \includegraphics[clip, trim=0.0in 0.0in 0.0in 0.73in, height=0.4\textheight]{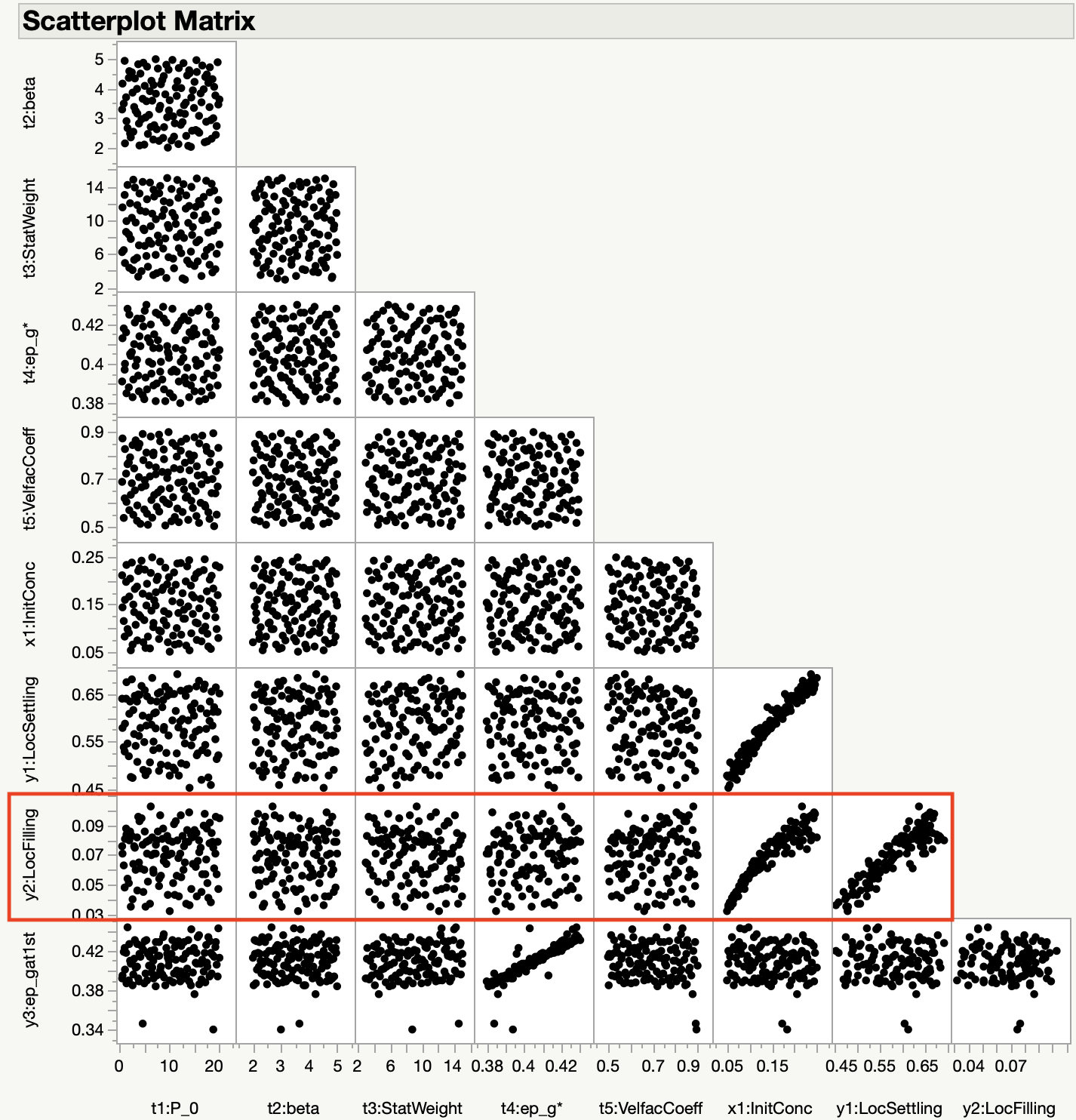}    
    \end{center}
    \caption{Scatter matrix plot of all input parameters and all retrieved output variables employed in the simulation campaign using OLH sampling (120 samples). Note that only the row highlighted in red rectangle is used in the study for the quantity of interest (i.e., y2:LocSettling).}
    \label{fig:C1scattermatrixALL}
\end{figure}

Similar to Figure~\ref{fig:SIM7_n60_i6_o3_scattermatrix}, the scatter matrix plot of the samples from simulation campaign \# 2 (only 120 samples) are shown in Figure~\ref{fig:C1scattermatrixALL}. 

Likewise, similar somewhat linear correspondences can be seen in the blocks ($t4$ v. $y3$), ($x1$ v. $y1$), and ($x1$ v. $y2$).

\begin{figure}[H]
    \begin{center}
    \includegraphics[clip, trim=0.0in 0.0in 0.0in 0.73in, height=0.4\textheight]{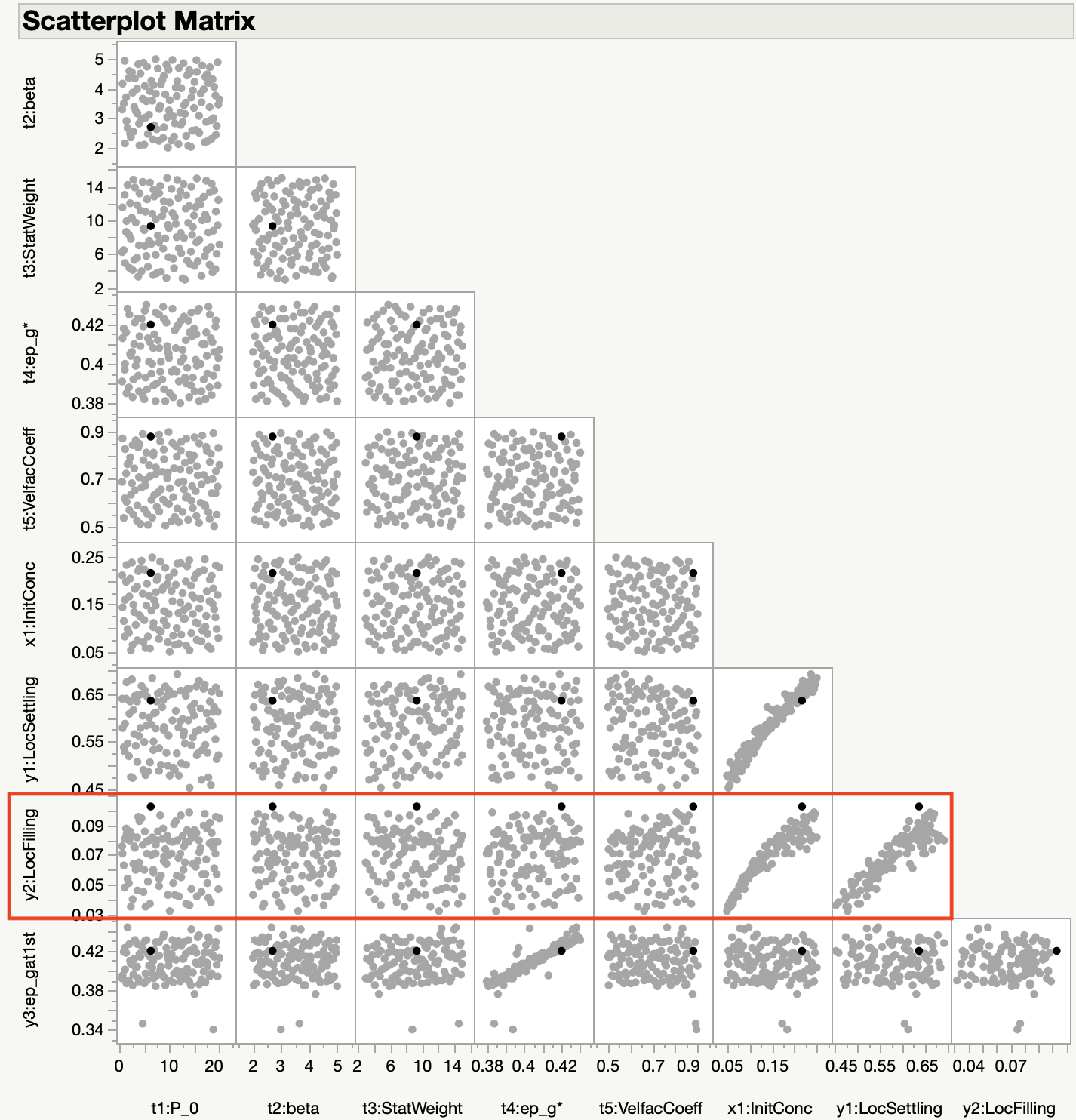} 
    \end{center}
    \caption{Scatter matrix plot of all input parameters and quantities of interest employed in simulation campaign using OLH design base (120 samples) with potential outlier sample example for y2:LocSettling (sample \# 118) highlighted in black color.}
    \label{fig:C1scattermatrixALLwOutlier}
\end{figure}

This type of qualitative visualization is also useful in identifying any apparent outliers among the quantities of interest. The identification of outliers can be done qualitatively with the help of visualizations as shown in Figure~\ref{fig:C1scattermatrixALL}, as these data points will appear apart from the majority of other data points such as sample \# 118, which is highlighted as a black colored circle in Figure~\ref{fig:C1scattermatrixALLwOutlier}. Outliers can be caused by non-converged simulations or unique input settings that create an extreme result for the quantity of interest. Hence, it is recommended to investigate the root cause for the results resembling an outlier and eliminate, if justified, as the outlier samples adversely affect the quality of the surrogate model. In this case, upon further investigation sample \# 118 was verified and deemed to be not an outlier.

\begin{figure}[H]
    \begin{center}
    \includegraphics[clip, trim=0.0in 0.0in 0.0in 0.0in, height=0.4\textheight]{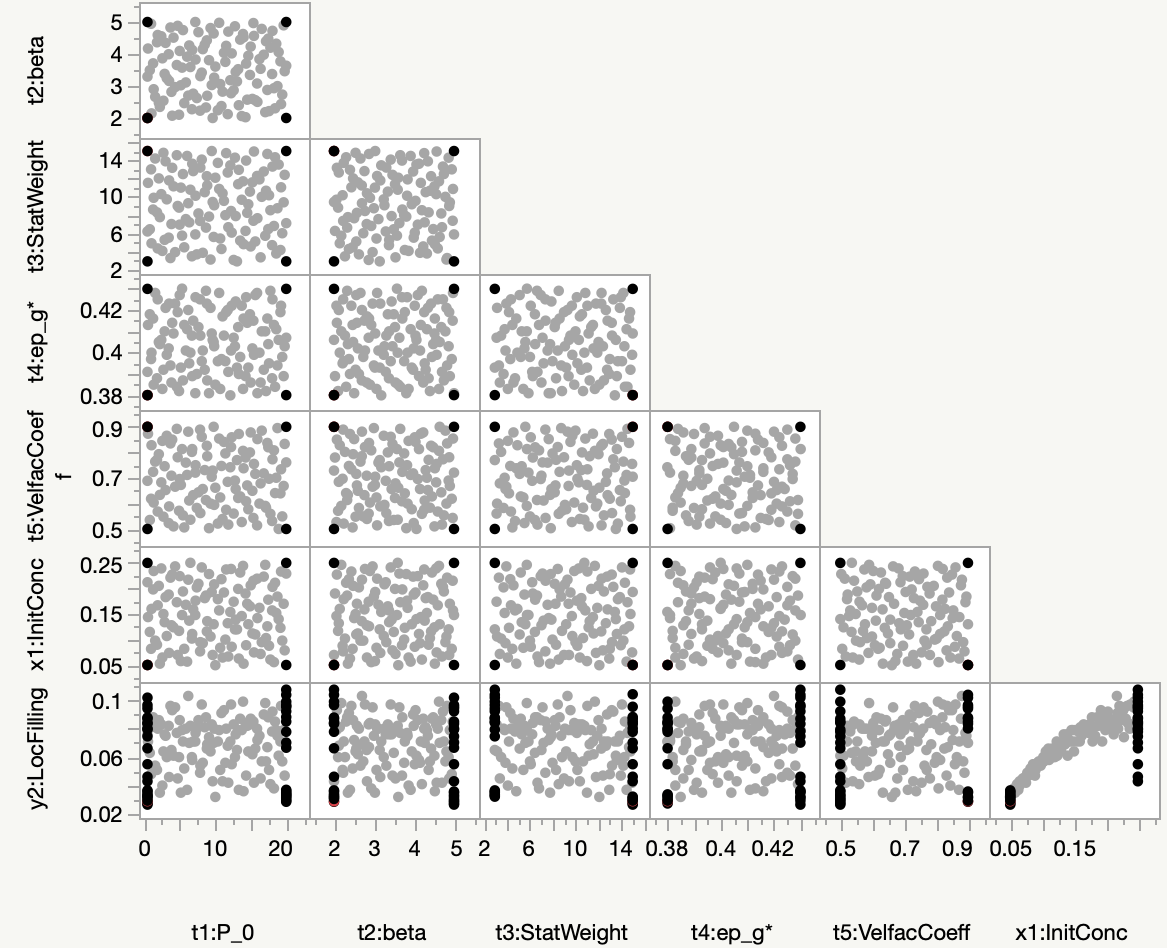}    
    \end{center}
    \caption{Scatter matrix plot of all input parameters and the selected quantity of interest employed in constructing a surrogate model using  all of the 184 samples with the new samples augmented shown in black colored circles.}
    \label{fig:C1scattermatrixALLwAugmented}
\end{figure}

The initial simulation campaign with 120 samples was augmented with 64 additional samples based on factorial sampling to ensure adequate samples existed on the boundary edges. Sampling methods such as Latin Hypercube attempt to satisfy the space filling property without any assurance on having adequate samples on the boundaries of the parameter space to be explored. Hence, an augmented sampling was employed to address this issue.
Figure~\ref{fig:C1scattermatrixALLwAugmented} shows the additional 64 samples (highlighted as black colored filled circles) on top of the original simulation campaign.   For the sake of brevity, only the selected quantity of interest (i.e., $y_2$:Location of Filling Shock) is shown in addition to the input parameters in this figure. The need for the additional samples to augment the initial 120 sample was identified after constructing a surrogate model with 120 samples.

\subsubsection{Surrogate Model Construction:} 

The next step in the workflow is to construct a surrogate model from the simulation campaign results, which will be used in lieu of the MFiX-PIC simulations for cheaper evaluations of the quantity of interest when performing Bayesian calibration study, in which GPM-based surrogate models are preferred due to the inherent ability to express the uncertainty. Hence, GPM surrogate model (option 10) in PSUADE was used.  

\begin{figure}[htp]
    \begin{center}
    \includegraphics[clip, trim=0.0in 0.125in 0.0in 0.1in, height=0.3\textheight]{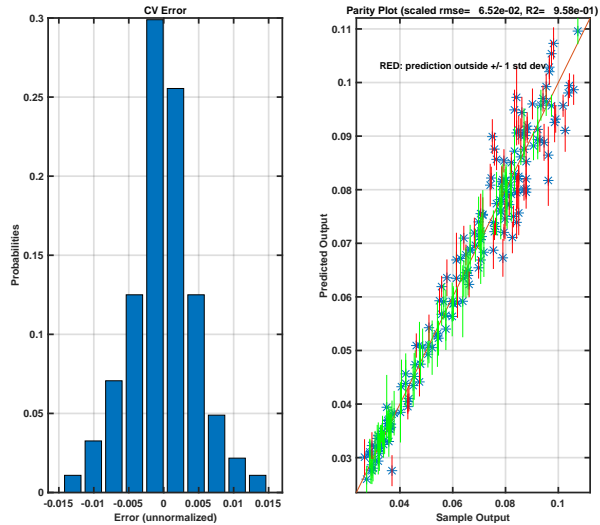} 
    \end{center}
    \caption{Assessment of the quality of the surrogate model (based on 184 samples) through cross-validation errors. }
    \label{fig:C2_n184_RSFA_cv_err}
\end{figure}

Similar to the assessment performed for the quality of surrogate model constructed for Simulation Campaign \# 1 (Figure~\ref{fig:C1_n60_RSFA_cv_err}), the 
 cross-validation error assessment results obtained for the surrogate model constructed with Simulation Campaign \# 2 results are shown in Figure~\ref{fig:C2_n184_RSFA_cv_err}. The parity plot on the right compares actual MFiX-PIC simulation results (horizontal axis with ``Sample Output'' label) with the surrogate model's predictions (vertical axis with ``Predicted Output'' label). Ideally, sample points (shown as blue asterisk symbols) should fall along a $45^{\circ}$ line through the parity plot.  Any deviation from the diagonal line reveals discrepancy between the simulation and data-fitted surrogate model, which implies an additional level of uncertainty being introduced when the surrogate model is used in lieu of the corresponding MFiX-PIC simulations. To better illustrate the error between simulation and data-fitted surrogate model, the histogram on the left reveals how the deviation from the diagonal is distributed. Ideal distribution of errors is expected to be centered around zero, have a narrow spread and tails without any skew.

\subsubsection{Global Sensitivity Analysis:}
A global sensitivity analysis was performed using the surrogate model constructed in the previous section to perform the necessary QoI evaluations for Sobols' Indices method in the PSUADE UQ Toolkit prior to the calibration study to better understand the most influential input parameters. In addition to five model parameters, a physical design variable $x_1$:Initial Concentration is considered for the sensitivity analysis.
For the sake of brevity, only results from simulation campaign \#2 were considered for sensitivity analysis due to large size of samples, which is expected to better represent the behavior of MFiX-PIC for given set of input and outputs.

\begin{figure}[htp]
    \begin{center}
    \includegraphics[clip, trim=3.in 2.6in 0.8in 2.6in, height=0.3\textheight]{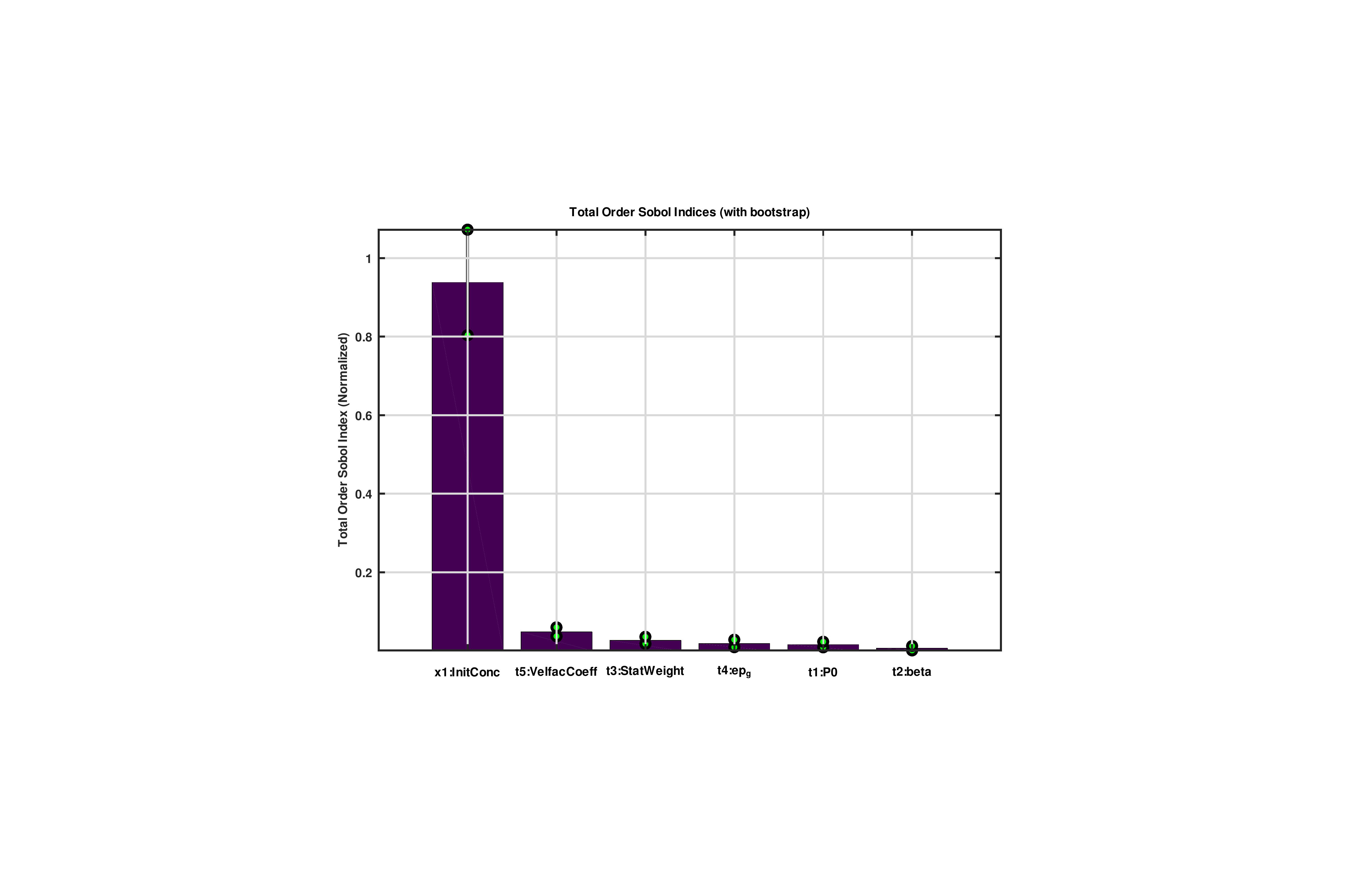}
    \end{center}
    \caption{Global sensitivity analysis results leveraging the surrogate model constructed from the 184 samples (simulation campaign \# 2) dataset.}
    \label{fig:C1_R2_rssoboltsi}
\end{figure}

Figure~\ref{fig:C1_R2_rssoboltsi} shows the Sobols'  Total Sensitivity analysis results which is aimed to identify quantitatively the most influential parameters on the quantity of interest, $y_2$:Location of Filling Shock. It is important to note that Total Indices take into account both main effects (such as $t1,t2,t3,t4,t5$ etc.) and their interaction effects on the quantity of interest.
For the 184 sample augmented simulation campaign results (i.e., the original simulation campaign with 120 samples augmented with 64 additional samples), $x_1$:Initial Concentration appears to have the most pronounced effect on $y_2$. The most influential second and third model input parameter were $t5$:VelfacCoeff and $t3$:StatWeight, respectively. The green symbols show the confidence interval associated with 300 sample bootstrapping for each parameter.  Confidence intervals do not show significant variability for any of the estimated Sobol' indices.

\subsubsection{Deterministic Calibration with PSUADE UQ Toolkit:}

Although a detailed discussion of the ``Deterministic Calibration'' method within the PSUADE UQ toolkit was presented in~\cite{Gel2021TRS}, a similar discussion is presented here for the reader as a reminder of the process, prior to discussion of Bayesian calibration.
Table~\ref{tab:C1_PropSettings} shows one of the proposed settings for the five modeling parameters obtained at the end of the deterministic calibration procedure.  Recall that this process involved deterministic optimization which finds the values of $\bm{\theta} : \{\theta_1 \dots \theta_5\}$ that minimize the residuals shown in Eqn. \ref{eqn:minimizeRes}.

\vspace*{1.2mm} 

There are various optimization techniques that may be employed to solve the residual minimization problem. In this case, the constructed data-fitted surrogate model is used to evaluate the $S_i(\bm{\theta},\bm{x})$ term instead of running MFiX-PIC simulations.

The deterministic calibration results presented in Table~\ref{tab:C1_PropSettings} were obtained with PSUADE UQ software \citep{PSUADE} through the following steps using simulation campaign \# 1 dataset:
\begin{enumerate}
\item Post-process and import the simulation campaign results into a format that PSUADE can read, i.e., standard ASCII text file with tabulated data where each column represents the input parameters considered and the quantity of interest. For this case, six columns of input ($\theta_1 \dots \theta_5$, and $x_1$) and one column of quantity of interest ($y2:\text{Location of Filling Shock}$) were employed. For formatting purposes, the first row of the file indicates total number of samples, total number of input parameters, and total number of quantities of interest.

\item Construct a data-fitted surrogate model in PSUADE to characterize the relationship between input parameters 
and quantity of interest, which in turn will be used for quick function evaluations needed during the optimization process. To minimize the effect of surrogate model related uncertainties, test the adequacy of the constructed surrogate model by employing cross-validation error assessment and other statistical measures such as $R^2$ if employing a polynomial regression-based surrogate model. The goal is to find the best suited data-fitted surrogate model for the given dataset.
\item Export and compile the constructed surrogate model as a standalone executable code (where PSUADE offers C and Python choices).  The executable code will then be used to perform function evaluations; passing in settings of $\theta_1 \dots \theta_5$ will return the quantity of interest as a scalar value.
\item Modify the C code for the exported surrogate model. The reason for this modification is that exported C code is structured to perform function evaluations, i.e., accept input and compute the quantity of interest (i.e., filling shock location) as output. However, the optimization procedure used in deterministic calibration aims to find the set of model parameters that minimize the residual. That means the exported surrogate model code is modified to not only evaluate the quantity of interest but also to calculate the residual (Eqn.~\ref{eqn:minimizeRes}) by taking the difference of computed value ($S_i(\bm{\theta},\bm{x})$) and the corresponding experiment's quantity of interest ($y_i$). If this modification is not performed, the optimization will be attempted for the wrong objective.

\item Utilize Bound Optimization By Quadratic Approximation (BOBYQA) 
optimizer in PSUADE to perform an optimization to find the best set of $\theta_1 \dots \theta_5$ values that give the least residual. 
To explore all possible solutions, perform the optimization 10 or 15 times then assess which set of proposed settings are most common among the trials.
\end{enumerate}

Table~\ref{tab:C1_PropSettings} shows the proposed settings for the five model parameters as a result of the deterministic calibration study.  As discussed earlier, experimental results (in this case analytical solution) are used to guide the calibration study. Two set of results were obtained based on 11 and 21 samples of analytical solutions, respectively.
\begin{table}[htp]
\begin{center}
\includegraphics[clip, trim=0.0in 0.0in 0.0in 0.0in, height=0.2\textheight]{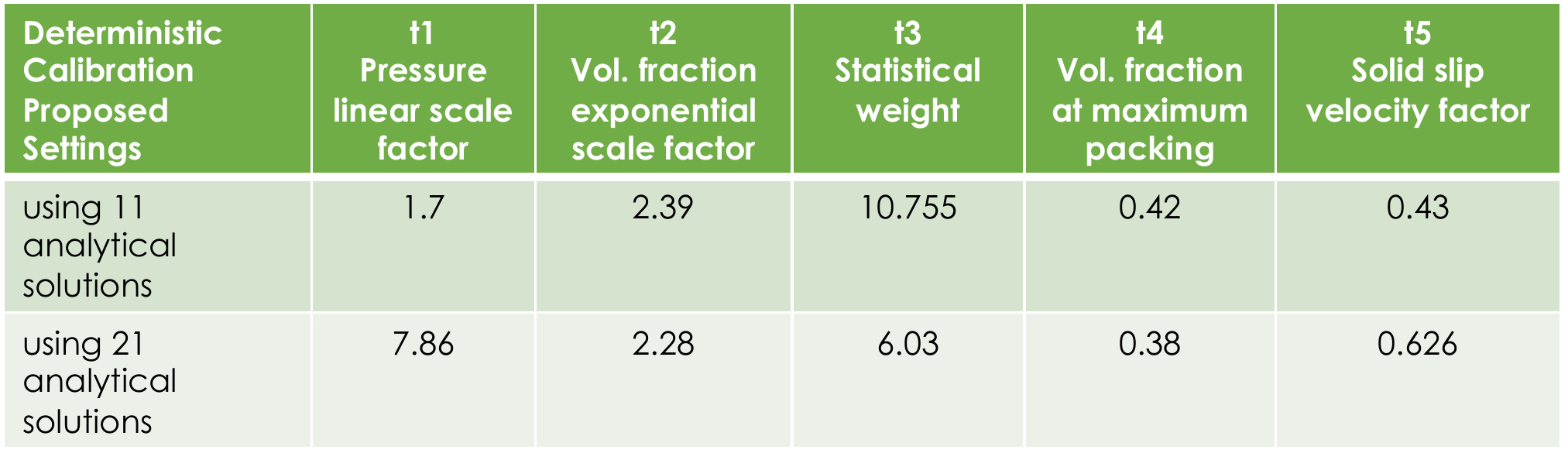}
 \caption{Proposed settings for the modeling parameters obtained through deterministic calibration using 11 versus 21 analytical solution samples to guide the calibration.}
    \label{tab:C1_PropSettings}
\end{center}
\end{table}
As mentioned, the values in Table~\ref{tab:C1_PropSettings} disagree with previously published data related to deterministic calibration in a settling bed using MFiX-PIC \citep{Gel2021TRS} due to a bug-fix in MFiX-PIC that was found and corrected with MFiX Release 21.2~\citep{MFIXBugFix21_2Release}.  Specifically, in prior simulations, inconsistent values for gas density were used within the PIC-modeling routines.  This bug was corrected and the simulation campaigns had to be carried out again, resulting in a new set of proposed values for the modeling parameters from deterministic calibration.

\subsubsection{Bayesian Calibration with PSUADE UQ Toolkit:}

The first few steps of the workflow presented in the previous section are also applicable for Bayesian calibration. Hence, a simulation campaign is designed and carried out to create the dataset for a data-fitted surrogate model that can ``adequately'' characterize the relationship between input parameters and the quantities of interest. Once the simulation campaign results are processed and the training dataset is constructed, a data-fitted surrogate model based on Gaussian Processes is employed. In the deterministic calibration, various surrogate model options were assessed to identify a best data-fitted surrogate model. In Bayesian calibration, usually a Gaussian Process model is preferred as it also provides uncertainty estimates for the surrogate model introduced.

The Bayesian calibration results presented in Table~\ref{tab:C1_PropSettingsBayesian} were obtained with PSUADE UQ software \citep{PSUADE} through the following steps:
\begin{enumerate}
\item Post-process and import the simulation campaign results into a format that PSUADE can read, i.e., standard ASCII text file with tabulated data where each column represents the input parameters considered and the quantity of interest. For this case, 6 columns of input ($\theta_1 \dots \theta_5$, and $x_1$) and 1 column of quantity of interest ($y2:\text{Location of Filling Shock}$) were employed. For formatting purposes, the first row of the file indicates total number of samples, total number of input parameters and total number of quantities of interest.

\item Construct a data-fitted surrogate model in PSUADE to characterize the relationship between input parameters 
and quantity of interest, which in turn will be used for quick function evaluations needed during the optimization process. For this step, GPM-based surrogate models are employed due to several additional features they offer for Bayesian calibration such as uncertainty of the constructed surrogate model.
To minimize the effect of surrogate model related uncertainties, test the adequacy of the constructed surrogate model by employing cross-validation error assessment. The goal is to ensure that an adequate data-fitted surrogate model can be constructed with the available training data-set. If necessary, additional sampling simulations may be carried out to augment the existing simulation campaign for better data-fitted surrogate model construction.
\item Choose the response-surface based MCMC simulation method for the estimate calculation of the posterior distributions for the model parameters. PSUADE offers two options: 1. Gibbs, which is slower, but it can be a more accurate method; 2. Brute-force approach, which is faster but accuracy-limited by sample size.
In addition, PSUADE requires several additional options that must be determined as part of this task, such as inclusion of surrogate model uncertainty and model discrepancy assessment. Also, several parameters for MCMC simulations need to be set, such as number of MCMC chains, maximum inference sample size, sample size to construct proposal distribution, etc. MCMC simulations are quite compute-intensive and advanced algorithms that necessitate careful consideration for proper setup. The reader is referred to PSUADE documentation~\citep{PSUADE} for additional details.
\item Evaluate the results of MCMC simulations using the PSUADE output and Matlab-based {\tt matlabmcmc2.m} file generated for successful completion of the MCMC. Figure~\ref{fig:C1MCMCposterior} shows the graphical output generated at the end of MCMC and saved in {\tt matlabmcmc2.m} when 11 samples of analytical solution are used to guide the calibration. Lower half of the figure shows the prior joint distributions (histograms within the blue colored triangle), the distributions in the upper half shows posterior joint distributions generated at the end of MCMC (histograms within the red colored triangle). The histograms on the diagonal show the marginalized posterior distributions, which is used to determine the proposed settings.
\begin{figure}[htp]
    \begin{center}
    \includegraphics[clip, trim=0.0in 0.0in 0.0in 0.0in, height=0.35\textheight]{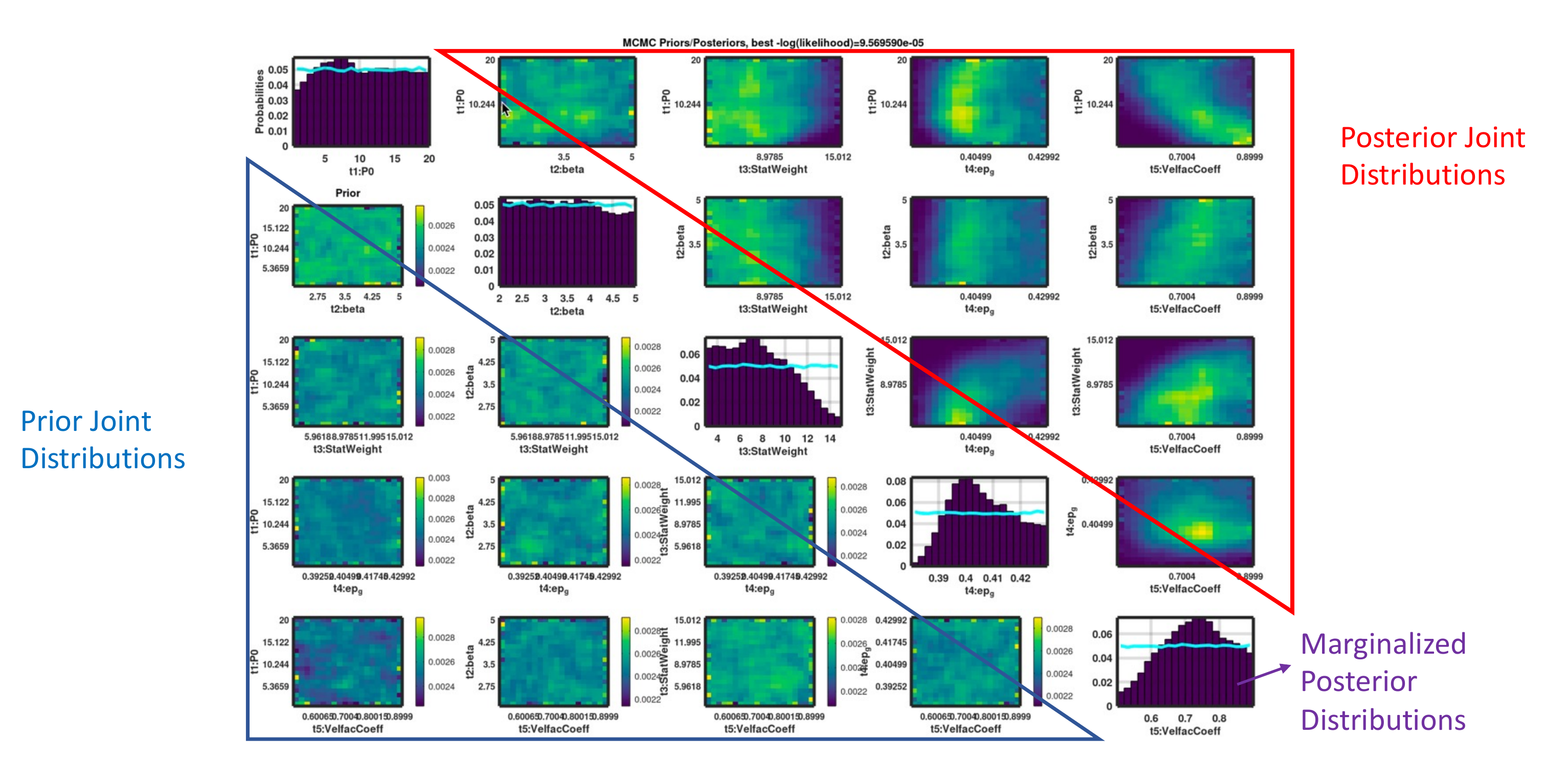}
    \end{center}
    \caption{Prior and posterior distributions generated. }
    \label{fig:C1MCMCposterior}
\end{figure}

\end{enumerate}

Table~\ref{tab:C1_PropSettingsBayesian} shows the proposed settings for the five model parameters as a result of the Bayesian calibration study.  As discussed earlier, experimental results (in this case, the analytical solution) are used to guide the calibration study. Two sets of results were obtained based on 11 and 21 samples of analytical solutions, respectively. When the actual values of the proposed settings are compared for each model input parameter in Table~\ref{tab:C1_PropSettingsBayesian}, it can be seen clearly that sample size of the analytical solution does not appear to demonstrate significant influence as both cases offered relatively similar proposed settings.
\begin{table}[htp]
\begin{center}
\includegraphics[clip, trim=0.0in 0.0in 0.0in 0.0in, height=0.15\textheight]{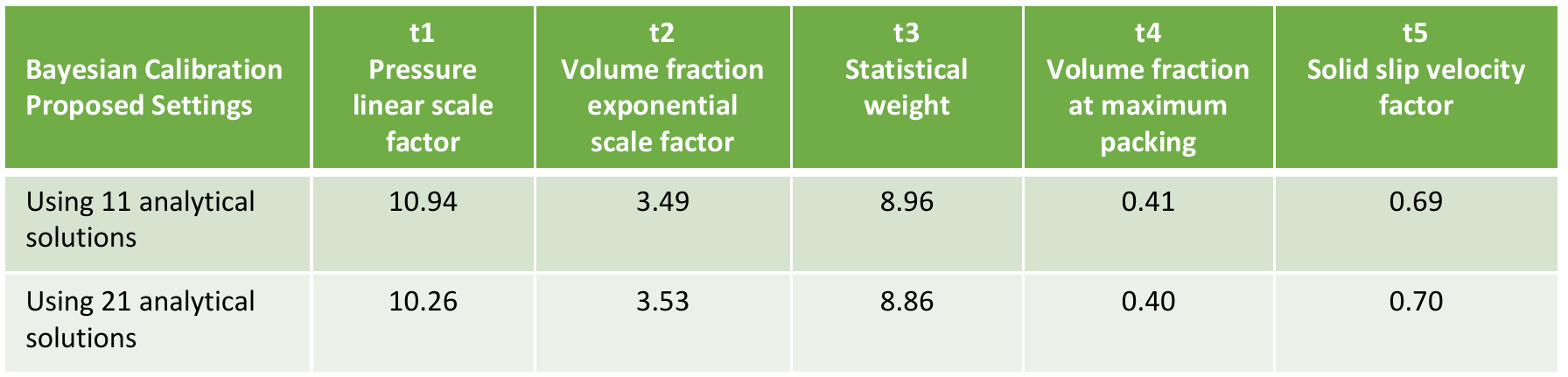}
 \caption{Proposed settings for the modeling parameters obtained through Bayesian calibration using 11 versus 21 analytical solution samples to guide the calibration.}
    \label{tab:C1_PropSettingsBayesian}
\end{center}
\end{table}

\paragraph{\bf Summary of the Proposed Settings from Deterministic \& Bayesian Calibration Studies:} 
Table~\ref{tab:C1_Comp_ALLFinal_brief} shows a summary of the proposed settings from the Deterministic (fourth and fifth columns from left), and Bayesian (sixth and seventh columns) calibration studies in addition to the default settings (second column), and V\&V Manual-based proposed settings (third column) for the same problem. In each calibration study, two separate proposed settings are shown, which correspond to 11 and 21 samples of analytical solutions, respectively. As discussed earlier, analytical solutions were used to guide the calibration study in lieu of actual experiments. To assess the sensitivity of the results towards number of samples available to guide the calibration, we performed two cases; (i) with 11 samples of analytical solutions (shown under columns titled ``{\tt PS\_Exp\_n11}''), (ii) with 21 samples of analytical solutions (shown under columns titled ``{\tt PS\_Exp\_n21}'').

\begin{table}[htb]
    \begin{center}
    \includegraphics[clip, trim=0.0in 1.78in 0.0in 0.0in, height=0.35\textheight]{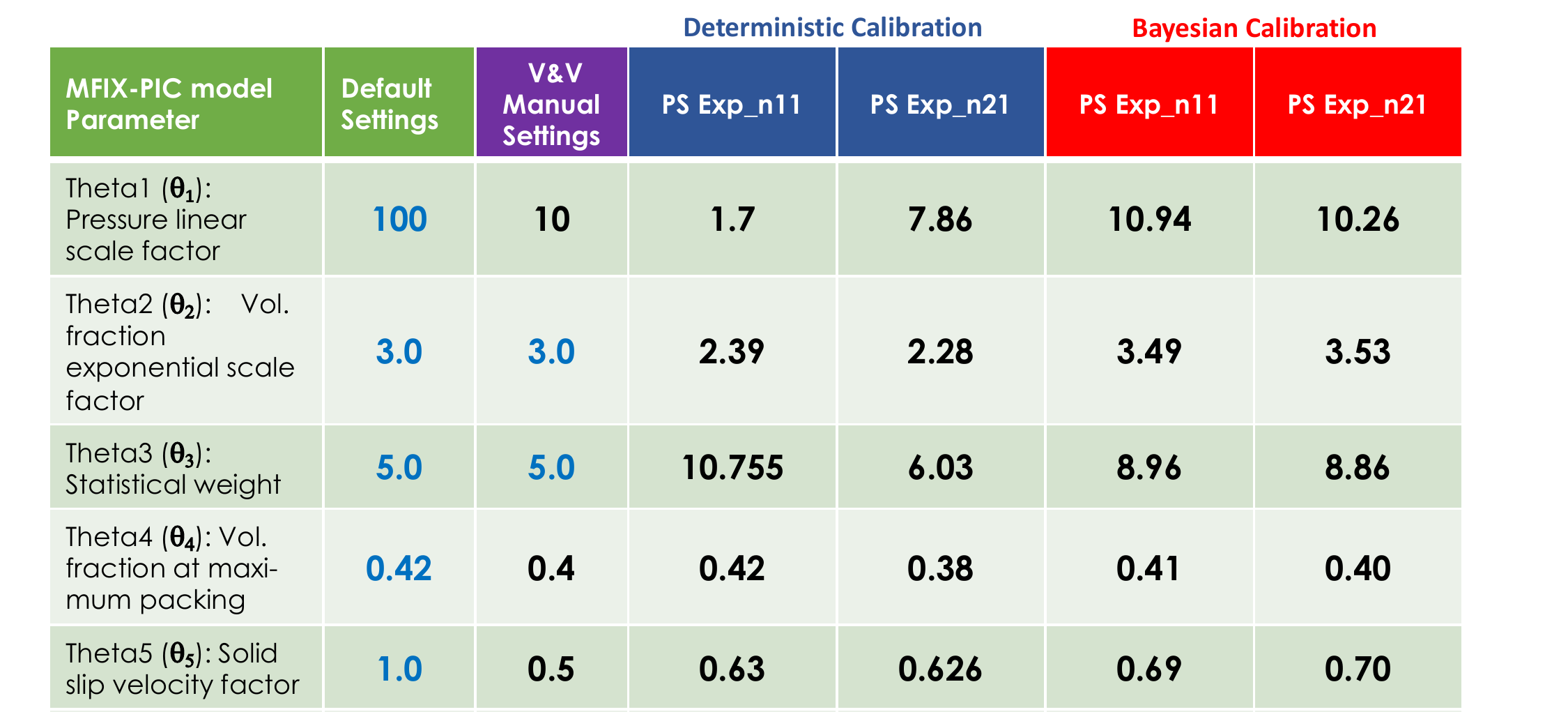}
    \end{center}
    \caption{Comparison of the proposed calibrated model parameters (based on PSUADE) with respect to default settings and V\&V Manual-proposed Settings.}
    \label{tab:C1_Comp_ALLFinal_brief}
\end{table}

In the comparisons performed, the Default Settings (second column in Table~\ref{tab:C1_Comp_ALLFinal_brief}) were determined by developers based on user experience and some prior literature-based settings. In fact, one of the major motivations of the current study was verifying if these values are sufficient to model the wide range of flow dynamics seen in industrial applications. 
In similar consideration, settings from the V\&V Manual~\citep{MFIXPICVV} were also used.

\paragraph{\bf Validation of the Accuracy Improvements Achieved with the Proposed Calibrated Settings:}  

Once a set of proposed settings for the five model parameters were obtained with both Deterministic and Bayesian calibration studies, the next step was to assess the accuracy improvement for the QoI that the proposed settings could offer.  It is important to note that the proposed calibrated model parameters potentially involve many sources of uncertainty.  These uncertainties originate from model errors propagated through the surrogate model (modeling assumptions, simplifications, and approximations included with the introduction of a surrogate model instead of actual MFiX-PIC simulations) and data provided as input. For this case, no experimental error exists since an analytical solution provided the comparative data set, but this is not the general case; care should always be taken in the verification process.

In order to assess the effectiveness of the proposed calibrated settings, a rigorous assessment based on 119 new samples was employed. 
The new samples were generated based on the fifth design variable, i.e., $x1\text{:Initial Concentration}$ settings in the range of 0.05 to 0.25, and were checked to ensure these were totally unseen samples, i.e.,  the same $x1$ setting was not used in the previous 120 sample campaign used to build the surrogate model.
Hence, a new  simulation campaign was carried out based on the 119 samples of $x1$ using the proposed calibrated model settings for each scenario outlined below as input to MFiX-PIC. The quantities of interest derived from the simulation results were then compared against the analytical solution obtained for each of the corresponding 119 samples using Eqn. \ref{eq:PercentRelError}.
\vspace{-3mm}
\begin{equation}
 \text{\% Relative Error} = \frac{(\text{Simulation} - \text{Analytical Solution})}{\text{Analytical Solution}}
     \label{eq:PercentRelError}
\end{equation}
Table~\ref{tab:C1_Comp_ALLFinal} is an extended version of Table~\ref{tab:C1_Comp_ALLFinal_brief} where three additional rows were added to assess the accuracy of the proposed settings and to compare against each other.
The last three rows in the table show the \% Relative Error computed for the 119 samples as part of the unseen verification runs. For each set of proposed settings, an average \% Relative Error is computed by taking the  \% Relative Error calculated for each of the 119 samples and taking their average reported for comparison. Also, the maximum and minimum \% Relative Error are reported. One common observation from the reported \% Relative Error results is that all are showing under-prediction consistently, but the extent is different between different studies.

\begin{table}[H]
    \begin{center}
     \includegraphics[clip, trim=0.0in 0.3in 0.0in 0.0in, height=0.4\textheight]{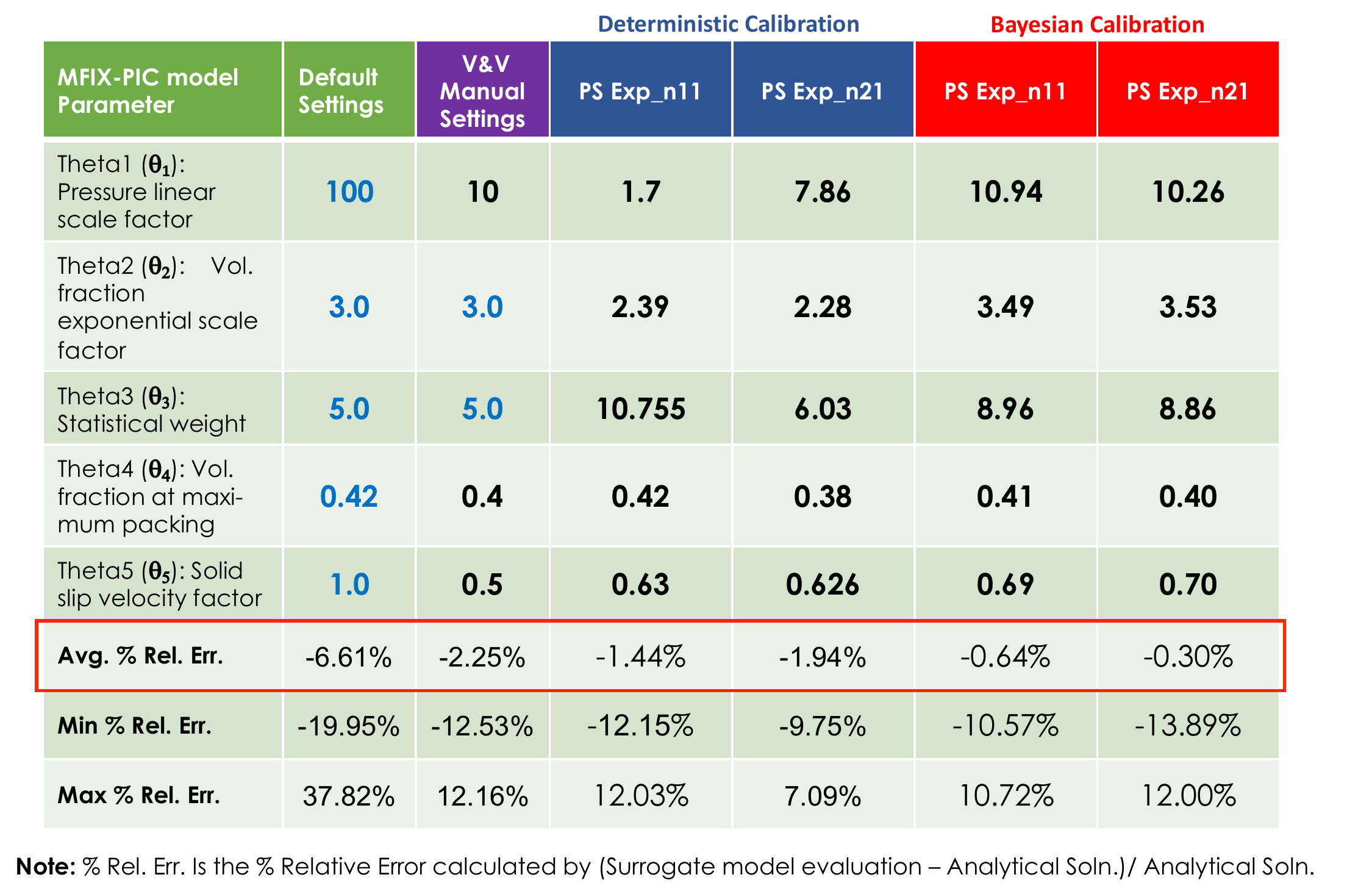}
    \end{center}
    \caption{Comparison of the \% Relative Error computed from proposed calibrated model parameters (based on PSUADE), default settings and V\&V Manual-proposed settings.}
    \label{tab:C1_Comp_ALLFinal}
\end{table}

As seen from the results presented in Table~\ref{tab:C1_Comp_ALLFinal}, when the reported Average \% Relative Error is considered (the row highlighted with a red rectangle), both calibration studies have substantially improved the accuracy of MFiX-PIC predictions compared to the default settings used in MFiX-PIC. In particular, Bayesian calibration has reduced the average \% Relative Error from -6.6\% to -0.3\% (i.e., calibration performed with 21 samples of analytical solution), which is nearly a twenty-fold improvement. Deterministic calibration results also show improvements on the order of four-fold error reduction when compared to the default settings. The third column shows the V\&V Manual-based proposed settings, which was a trial-and-error based calibration study performed and documented in \cite{MFIXPICVV}. The improvement achieved in average \% Relative Error when compared to V\&V Manual proposed settings appears to be less (i.e., - 2.25\% versus -1.44\% with Deterministic calibration), but still significant, especially for Bayesian calibration study results (i.e., -2.25\% versus -0.3\%).

\begin{figure}[H]
    \begin{center}
    \includegraphics[clip, trim=0.15in 0.0in 0.0in 0.0in, height=0.225\textheight]{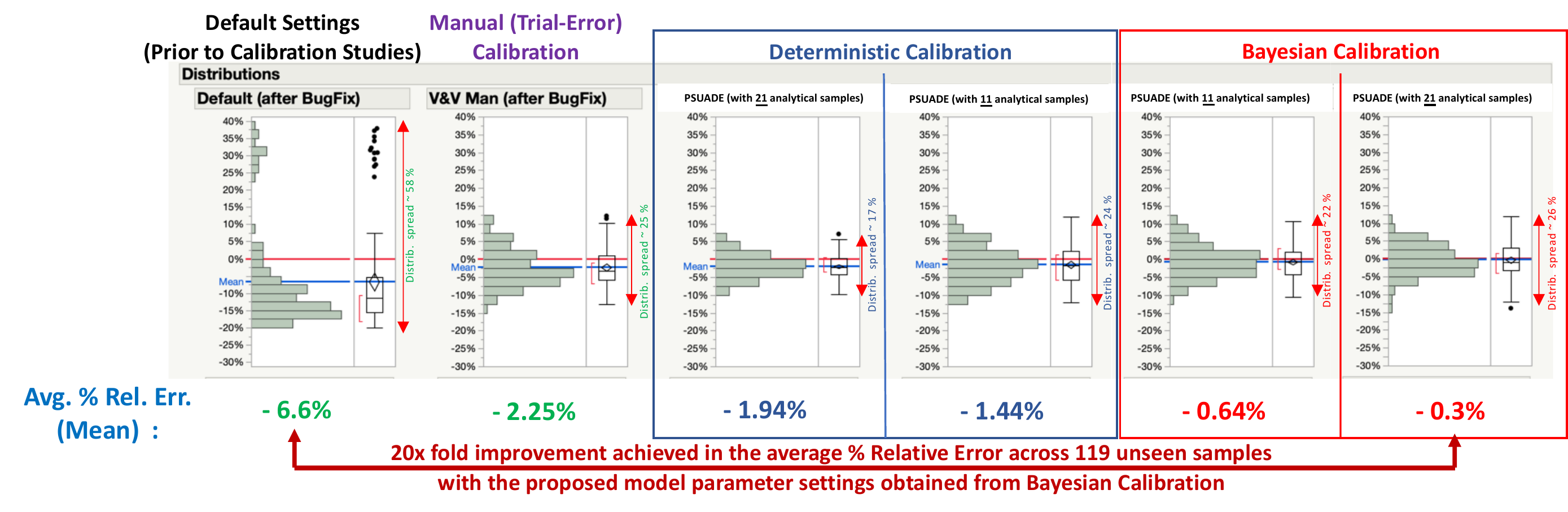}
    \end{center}
    \caption{Histogram Comparison of \% Relative Error for quantity of interest, y2: Location of Filling Shock before calibration (Default Settings) and after using Deterministic and Bayesian calibration.}
    \label{fig:C1_HistogramErrorComp_ALL}
\end{figure}

Although the average and max/min values of the \% Relative Error reported in Table~\ref{tab:C1_Comp_ALLFinal} are useful to make comparisons, how these errors are distributed was also investigated. Figure~\ref{fig:C1_HistogramErrorComp_ALL} shows the histograms for the \% Relative Error distributions calculated from the 119 unseen samples of verification runs for each set of proposed model parameter settings from Deterministic and Bayesian calibration and the default settings. Ideally, all of the errors from sampling simulations should be centered at 0.0\% with a narrow spread around the red horizontal line in each histogram. The blue horizontal line shows the average of the \% Relative Error from all 119 simulations used as part of the validation process for each set of proposed model parameters. Hence, the blue line closer to red line is preferred. The left most histogram shows the \% Relative Errors when default settings are used. The average of \% Relative Error is -6.6\%, which indicates under-prediction below the red horizontal line. The second histogram from the left shows V\&V Manual proposed settings obtained as trial-and-error and published in \cite{MFIXPICVV}, which appears to have better error distribution when compared to the histogram for the default settings.  Third and fourth histograms in Figure~\ref{fig:C1_HistogramErrorComp_ALL} show the \% Relative Error distribution obtained from Deterministic calibration when 21 and 11 analytical samples are used to guide the calibration, respectively. Similar, fifth and sixth histograms from the left show the results from Bayesian calibration, again using 11 and 21 samples. As seen from the histograms, calibration improved the distribution of errors by moving the average \% Relative Error from the 119 sample verification runs closer to the center at 0\% (red horizontal line). Likewise, calibration tightened the spread around the mean. In particular, the Bayesian calibration provided the best accuracy improvement by having the mean \% Relative Error at approximately 0.3\%. In the previous report \citep{Gel2021TRS}, a standalone section was dedicated to analyze the effect of number of samples available from experiments (in this case analytical solution) to guide the calibration. For the sake of brevity, in the current report, the effect of sample size on the calibration is assessed by direct comparison of 11 versus 21 samples of analytical solution side by side. Having more data points to guide the calibration improved the Bayesian calibration results (i.e., 11 analytical solution-based results yielded -0.64\% average relative error whereas 21 analytical solution-based results yielded -0.3\%). However, for the deterministic calibration the opposite effect was observed. When we investigated, the plausible cause was associated with the formulation of the deterministic calibration which attempts to re-frame the problem as a residual minimization problem and its optimization. Hence, the minimization search most likely hit a local minima in the case of 21 samples, making it more difficult to satisfy convergence criteria, as compared to an 11 sample-based minimization. In spite of this finding, both calibration studies yielded substantially improved accuracy when compared with the default settings-based results.

\hypertarget{conclusions}{\section{Conclusions}\label{conclusions}}

The study presented in this report evolved from systematic Verification, Validation, and Uncertainty Quantification efforts initiated at the U.S. Department of Energy's (DOE) NETL since 2010 to assess the credibility of the MFiX suite of solvers~\cite{MFIX}. This report follows three earlier related reports: (1) Verification and Validation of MFiX-PIC~\cite{MFIXPICVV}, (2) Sensitivity Analysis of MFiX-PIC,~\cite{Vaidheeswaran2020}, (3) Deterministic Calibration of MFiX-PIC, Part 1: Settling Bed,~\cite{Gel2021TRS} .
The first study aimed to capture and document any discrepancy noted in MFiX-PIC by comparing simulation results to available experimental data directly. The second study aimed to evaluate the sensitivity of keyword-accessible modeling input parameters employed in MFiX-PIC, by examining several quantities of interest as modeling input parameters varied. The third study was the first demonstration of a systematic calibration (in particular, deterministic) study applied to a problem configuration of interest. The current study is a follow-up to the third report where another calibration methodology is employed for the same problem and the accuracy improvements from both approaches are directly compared.

The primary focus of this study was to develop and demonstrate a procedural methodology for improving the credibility of MFiX-PIC simulations by applying both Deterministic and Bayesian calibration methods to the modeling parameters considered. Initially, three targeted applications were selected to encompass widely encountered flow configurations: particle settling, fluidized bed and circulating fluidized bed. This report documents the Bayesian calibration of five MFiX-PIC modeling parameters in the context of the first targeted application, particle settling. A unique advantage for this problem is an analytical solution for the quantity of interest, location of the filling shock. 
This enabled the calibration study to proceed without any physical experiments. In addition to the Bayesian calibration study, the findings of the Deterministic calibration study from ~\cite{Gel2021TRS} were also considered for direct comparison to assess which calibration method improved MFiX-PIC accuracy for this particular application.

Two separate datasets with 11 and 21 samples based on the analytical solution for the location of the filling shock were generated by varying the initial solids concentration parameter ($x_1$) within the interval $[0.05,0.25]$. Then these data-sets were used similar observations from an experiment. Although, no experimental uncertainty was considered for deterministic calibration, some artificial uncertainty was introduced in the Bayesian calibration study to mimic an actual experiment-based calibration study.

The deterministic calibration procedure can be framed as the minimization of residuals (simulation - experiment) problem. In order to perform the thousands of evaluations required while testing different model parameter settings during optimization, a data-fitted surrogate model was constructed.  This model, after assessing its quality for characterizing the relationship between input and output data-sets, was then used instead of actual MFiX-PIC simulations to save time.  Note that running thousands of MFiX-PIC simulations as part of an optimization loop would be prohibitively time-consuming.

The Bayesian calibration procedure is a more advanced approach that is established on a statistical framework in which a prior distribution for each model parameter targeted for calibration is assumed, which can be based on prior data, literature or expert opinion. Their posterior distribution is obtained via a MCMC-based approach, which is computationally expensive. In lieu of actual MFiX-PIC simulations to carry out MCMC, a data-fitted surrogate model based on Gaussian Processes is utilized, which requires an initial simulation campaign to ensure an adequate quality surrogate model can be constructed to characterize the relationships between input parameters and quantities of interest. Bayes' Theorem is utilized to update the prior assumed distributions of the model parameters as part of the calibration process. The result of Bayesian calibration is a posterior distribution for each model parameter instead of a single scalar value. Hence, the uncertainty associated with the model parameters and their effect can be more effectively assessed. 

Due to the long time span of the studies and a bug discovered in the MFiX-PIC causing calculation-based inconsistency between {\tt ROP\_g} (product of gas density times gas volume fraction) and {\tt EP\_g} (gas volume fraction) necessitated multiple simulation campaigns with different sample sizes to be carried out and then re-run again due to the bug fix implemented. Initially, a simulation campaign with 120 samples of actual MFiX-PIC simulations was designed using the OLH sampling method for the six input parameters considered. Based on the insight gained from the earlier simulation campaigns, a new campaign (referred as simulation campaign \# 1) with the half of the original sample size (i.e., 60 samples) was constructed with minor revisions in the lower and upper bounds for the six parameters, as shown in Table~\ref{tb:InputParams_OLHn60}. The settings used for the six input parameters and the quantity of interest acquired from the simulation campaign results were compiled in a tabular formatted ASCII file to be used as input to UQ toolkit software for constructing a surrogate model and for performing the remainder of the analysis. For the Bayesian calibration study, the original 120 samples-based simulation campaign (referred as simulation campaign \# 2) was employed. However, the simulation campaign was later augmented with 64 additional samples to capture the responses at the edges of the parameter space when initial Bayesian calibration results indicated inadequacy of the initial simulation campaign.

UQ software, PSUADE, from Lawrence Livermore National Laboratory~\citep{PSUADEweb} was employed for data-fitted surrogate model construction, surrogate model adequacy and quality checks.  In addition, PSUADE was also used for performing the minimization of residuals to obtain the best set of model parameter settings. 

The effectiveness of the proposed calibrated model parameter settings obtained as a result of the calibration processes were evaluated by running additional MFiX-PIC simulations using the new settings for model parameters ($\theta_1, \theta_2, \theta_3, \theta_4, \theta_5$) and then calculating \% relative error with respect to the analytical solution for the location of the filling shock at the corresponding $x_1$ settings. In addition, a new set of simulations was performed at the same $x_1$ settings using both the default MFiX-PIC settings and those proposed in the V\&V Manual for the five model parameters ($\theta_1, \theta_2, \theta_3, \theta_4, \theta_5$). An overall comparison of \% relative error from each simulation was presented as well as a histogram view of \% relative errors.
Note that the availability of an analytical solution enabled performing an error assessment for each sampling simulation. This type of precise error assessment is typically not feasible when physical experiments are utilized for the calibration process.  

The proposed calibrated model parameter settings obtained with PSUADE UQ software were superior to the default MFiX-PIC settings and the settings proposed in the V\&V Manual.    
The \% relative error histogram plots for the proposed calibrated model parameter settings were demonstrated to yield substantially more accurate MFiX-PIC results for the particle settling application. 
This was shown with a 
rigorous approach that utilized 119 unseen samples of $x_1$ settings. Again, both deterministic and Bayesian calibration-based proposed settings outperformed the other settings for the five model parameters as clearly seen in Table~\ref{tab:C1_Comp_ALLFinal} and Figure~\ref{fig:C1_HistogramErrorComp_ALL}.

\begin{table}[htp]
    \begin{center}
    \includegraphics[clip, trim=0.0in 0.0in 0.0in 0.0in, height=0.25\textheight]{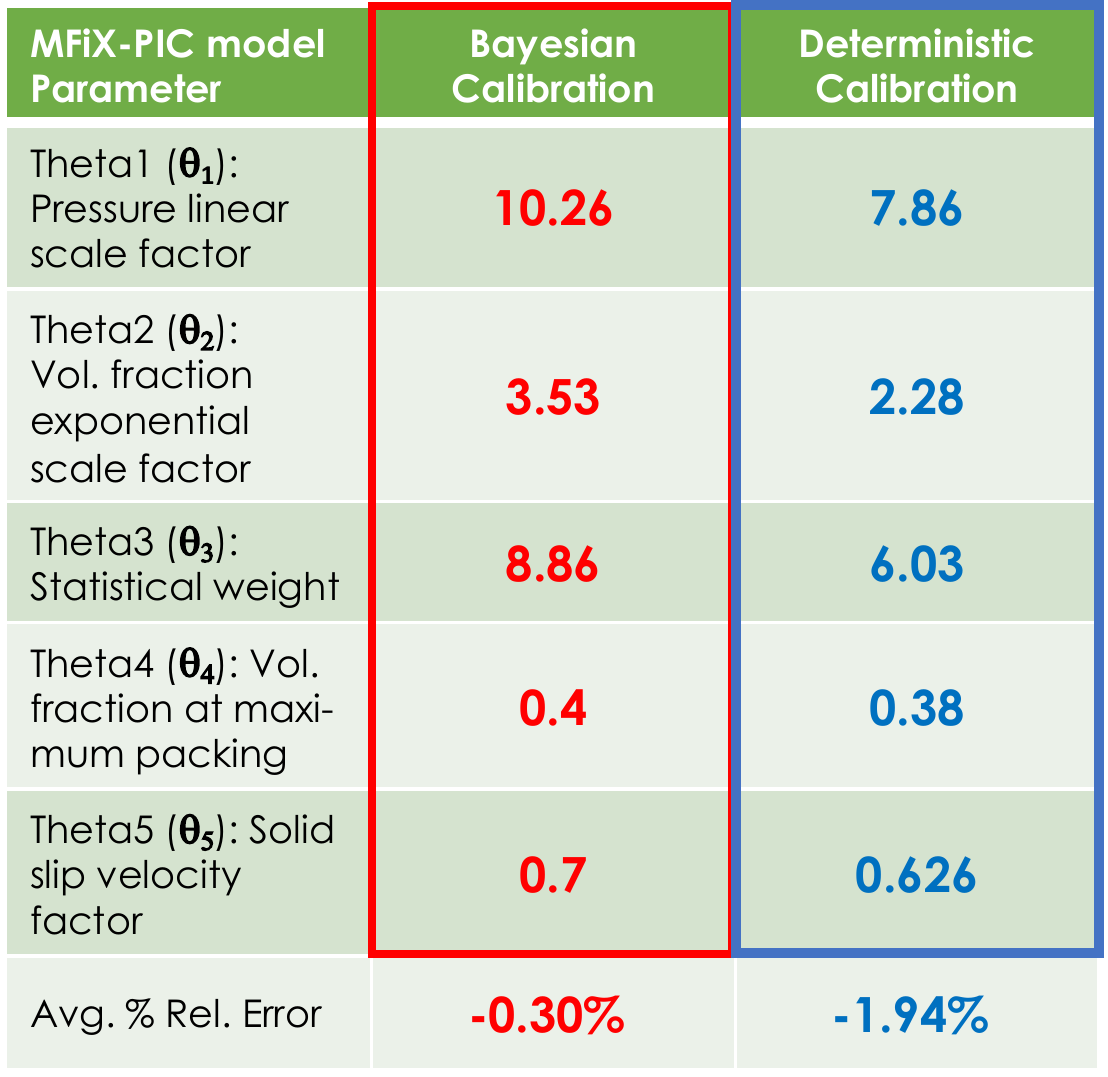}
    \end{center}
    \caption{Validated calibrated model parameters based on PSUADE and Nodeworks results.}
    \label{tab:C1_FinalProposedSettings}
\end{table}

\begin{figure}[H]
    \begin{center}
    \includegraphics[clip, trim=0.0in 0.0in 0.0in 0.0in, width=0.5\textwidth]{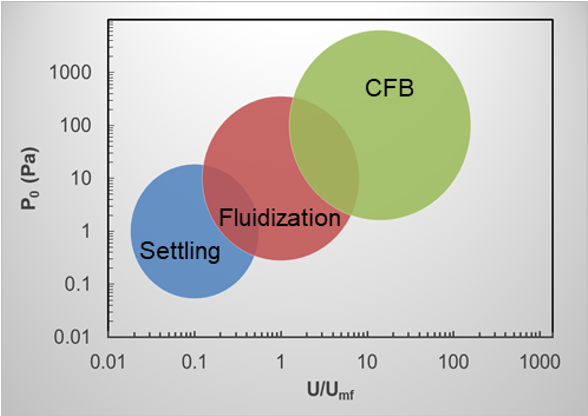}
    \end{center}
    \caption{Hypothetical flow regime map. \cite{Vaidheeswaran2020}}
    \label{fig:flowregimes}
\end{figure}

Finally, 
Table~\ref{tab:C1_FinalProposedSettings} 
shows the proposed calibrated settings for MFiX-PIC model parameters suggested for use in applications similar to the particle settling case. Although, both Bayesian and Deterministic calibration-based results have been demonstrated to give more accurate MFiX-PIC simulation results compared to default settings and V\&V Manual-proposed settings, Bayesian calibration appears to yield more accuracy increase for the particular case. The average \% relative error from Bayesian calibration and Deterministic calibration were -0.3\% and -1.94\%, respectively. Both demonstrated tight distribution of the errors around the mean.  The amount of effort and computational expense to achieve Bayesian calibration was substantially more than the Deterministic calibration. Hence, Bayesian calibration settings could be used for applications that fall within the particle settling region of the hypothetical flow regime map shown in Figure~\ref{fig:flowregimes} and are expected to give more accuracy for MFiX-PIC.\\

Several suggestions for future work are based on investigation of the effect of surrogate model quality on the calibration study results. For example, in addition to the simulation campaign \# 1 based deterministic calibration study, it is worthwhile to carry out the same study with 120 samples or 184 samples using the existing datasets without the need to run a new campaign. Also, all of the available datasets (60 + 120 + 64) can be combined together after screening for potential overlapping samples and removed to create a large dataset of simulation campaign output. Due to the time constraints, the authors could not complete this study, but the necessary input files are provided in the Appendix for those readers interested to carry out the study. Another potential future study could be better fine tuning of the Gaussian Process Models constructed as a surrogate model as the current study used the default settings in the UQ toolkit, which offers more advanced features to tune for a better data-fitted surrogate model fit. Finally, the particle settling case was unique in the sense that an analytical solution was available to be used in lieu of data to guide the calibration that is typically acquired through physical experiments. The analytical solution offered to create any number of datasets (e.g., 11 samples versus 21 samples). However, the effect of uncertainty in this dataset which is the typical in physical experiments was not investigated in detail. For Bayesian calibration certain level of artificial uncertainty was introduced, but the sensitivity of the calibration results to the uncertainty in the experimental (artificially introduced in the analytical solution for this case only) was not investigated. Hence, a systematic sensitivity study by testing the effect of different levels of uncertainty in the 11 and 21 sample analytical solution results could provide a better insight on this aspect, which should become more important for the following problem of fluidization where physical experiment dataset with replication measurements are available.

A separate dedicated report will demonstrate deterministic and statistical calibration for the second case, fluidized bed applications. The overall goal of these reports is to establish validated guidance for MFiX-PIC users who are planning to carry out simulations that fall within the hypothetical flow regimes explored while also offering a unified set of proposed calibrated settings.

\section*{Acknowledgments}
This work was performed in support of the U.S. Department of Energy's (DOE) Fossil Energy Crosscutting Technology Research Program. The research was executed through the National Energy Technology Laboratory's (NETL) Research and Innovation Center's CFD for Advanced Reactor Design (CARD) Field Work Proposal.

The authors wish to acknowledge Dr. William A. Rogers for programmatic guidance, direction,
and strong support to carry out this research. 

The authors would like to thank Dr. Charles Tong of Lawrence Livermore National Laboratory for the support he provided in using PSUADE, including enabling PSUADE to perform a deterministic calibration workflow by modifying the exported surrogate model and clarifications on the use of Bayesian Calibration features of PSUADE. Finally, the authors would like to acknowledge and thank Dr. Jean-Fran\c{c}ois Dietiker for his continuous feedback and suggestions to improve the quality of the study.

\section*{Disclaimer}
This project was funded by the U.S. Department of Energy, National Energy Technology Laboratory, in part, through a site support contract. Neither the United States Government nor any agency thereof, nor any of their employees, nor the support contractor, nor any of their employees, makes any warranty, express or implied, or assumes any legal liability or responsibility for the accuracy, completeness, or usefulness of any information, apparatus, product, or process disclosed, or represents that its use would not infringe privately owned rights.  Reference herein to any specific commercial product, process, or service by trade name, trademark, manufacturer, or otherwise does not necessarily constitute or imply its endorsement, recommendation, or favoring by the United States Government or any agency thereof. The views and opinions of authors expressed herein do not necessarily state or reflect those of the United States Government or any agency thereof.

\bibliographystyle{unsrtnat}
\bibliography{references}  

\newpage
\section*{Appendix}                                     
\addcontentsline{toc}{section}{APPENDIX}    
\label{AppHeader}
\renewcommand{\thesection}{\Alph{section}.\arabic{section}}

\appendix

The purpose of this Appendix is to provide information necessary for the reader to reproduce the results of this report as much as possible due to inherent randomness in the process and potential differences of the installed UQ software versions.  There is expectation that the reader already has software access, as well as the necessary advanced skill to work within and analyze results from associate software.  Also, the provided tabular input dataset can be used to test different methods as they represent the consolidated results from simulation campaigns with the provided details. For the readers interested in the missing aspects or other relevant questions, it is recommended to reach the lead author of the publication via email (aike@alpemi.com).

The files discussed in this section are available through NETL's Gitlab repository under the following URL: \\
\url{https://mfix.netl.doe.gov/gitlab/quality-assurance/PIC_calibration/-/tree/main/Case1_ParticleSettling}

Registration requirement to the NETL Gitlab repository has been removed and made public.

\noindent Users can clone the repository for all PIC Calibration related studies with the following {\tt git clone} command from a Linux console terminal, then navigate to the folder where Deterministic Calibration related files reside:

\begin{lstlisting}
> git clone https://mfix.netl.doe.gov/gitlab/quality-assurance/PIC_calibration.git
> cd PIC_calibration/Case1_ParticleSettling
\end{lstlisting}

For those who use a GUI based Git client, users can point to \url{https://mfix.netl.doe.gov/gitlab/quality-assurance/PIC_calibration.git} and clone the repository to their local system.

The repository consists of three top level subdirectories as shown below:
\begin{itemize}
\item \href{https://mfix.netl.doe.gov/gitlab/quality-assurance/PIC_calibration/-/blob/main/Case1_ParticleSettling/DeterministicCalibration}{\tt {\bf DeterministicCalibration}} contains the files employed for the deterministic calibration study that was previously published ~\citep{Gel2021TRS}. Please note the results in this folder are based on MFiX-PIC prior to the bug fix which was released with MFiX Release 21.2~\cite{MFIXBugFix21_2Release}.
\item  \href{https://mfix.netl.doe.gov/gitlab/quality-assurance/PIC_calibration/-/blob/main/Case1_ParticleSettling/DeterministicCalibration_UpdatedResultsDueBugFix}{\tt {\bf DeterministicCalibration\_UpdatedResultsDueBugFix}} contains the input files generated from simulation campaign \# 1 to perform the deterministic calibration study as documented in this report. Due to the above mentioned bug fix issue a separate directory was created for the deterministic calibration study while preserving the original results under  \href{https://mfix.netl.doe.gov/gitlab/quality-assurance/PIC_calibration/-/blob/main/Case1_ParticleSettling/DeterministicCalibration}{\tt DeterministicCalibration}.
\item  \href{https://mfix.netl.doe.gov/gitlab/quality-assurance/PIC_calibration/-/blob/main/Case1_ParticleSettling/BayesianCalibration}{\tt {\bf BayesianCalibration}} contains the input files generated from simulation campaign \# 2 to perform the Bayesian calibration study presented in this study.
\end{itemize}
A directory tree is shown in the file \href{https://mfix.netl.doe.gov/gitlab/quality-assurance/PIC_calibration/-/blob/main/Case1_ParticleSettling/README.md}{README.md} which provides an overview of the organization of the entire set of directories and stored files within this repository.
For the remainder of the Appendix, the operating system level command examples displayed assume the {\tt bash} shell environment. The reader should check their shell environment with ''{\tt echo \$SHELL}'' and make any necessary adjustments.

\section{Input Files Used for PSUADE}
\label{appendix:PS_RSM}

All PSUADE analyses were performed with version 2.0. It is assumed that the user has setup their environment and path to the PSUADE 2.0 binary properly.
\subsection{Importing External Dataset to PSUADE}
{\bf List of the files used with hyperlinks to the repository for \\ Simulation Campaign \# 1 (Deterministic Calibration): }\\
\href{https://mfix.netl.doe.gov/gitlab/quality-assurance/PIC_calibration/-/blob/main/Case1_ParticleSettling/DeterministicCalibration_UpdatedResultsDueBugFix/PSUADE/C1_SIM7_Results_OLH_n60_i6_o3.xlsx}{\tt C1\_SIM7\_Results\_OLH\_n60\_i6\_o3.xlsx} : Microsoft\textsuperscript{\textregistered} Excel file with the simulation dataset including input parameters and all QoIs retrieved from the simulation results.\\
\href{https://mfix.netl.doe.gov/gitlab/quality-assurance/PIC_calibration/-/blob/main/Case1_ParticleSettling/DeterministicCalibration_UpdatedResultsDueBugFix/PSUADE/C1_SIM7_OLH_n60_i6_o1_y2.dat}{\tt C1\_SIM7\_OLH\_n60\_i6\_o1\_y2.dat} : 60 samples with 6 input and 1 QoI (without column header titles) provided in plain ASCII data format\\
\href{https://mfix.netl.doe.gov/gitlab/quality-assurance/PIC_calibration/-/blob/main/Case1_ParticleSettling/DeterministicCalibration_UpdatedResultsDueBugFix/PSUADE/C1_SIM7_OLH_n60_i6_o1_y2.csv}{\tt C1\_SIM7\_OLH\_n60\_i6\_o1\_y2.csv} : 60 samples with 6 input and 1 QoI (including the column header titles) provided in CSV format\\
\href{https://mfix.netl.doe.gov/gitlab/quality-assurance/PIC_calibration/-/blob/main/Case1_ParticleSettling/DeterministicCalibration_UpdatedResultsDueBugFix/PSUADE/psData_SIM7_n60_i6_o1_y2}{\tt psData\_SIM7\_n60\_i6\_o1\_y2} : PSUADE native datafile generated after importing above file CSV or DAT file. Note that only y2:Location of Filling Shock is included as QoI. \\

{\bf List of the files used with hyperlinks to the repository for \\ Simulation Campaign \# 2: (Bayesian Calibration):  }\\
\href{https://mfix.netl.doe.gov/gitlab/quality-assurance/PIC_calibration/-/blob/main/Case1_ParticleSettling/BayesianCalibration/PSUADE/C2_SIM6_Results_OLH_n120_i6_o3.xlsx}{\tt 
C2\_SIM6\_Results\_OLH\_n120\_i6\_o3.xlsx} : Microsoft\textsuperscript{\textregistered} Excel file with the simulation dataset\\
\href{https://mfix.netl.doe.gov/gitlab/quality-assurance/PIC_calibration/-/blob/main/Case1_ParticleSettling/BayesianCalibration/PSUADE/C2_SIM6Merged_OLH_n184_i6_o1_y2.csv}{\tt 
C2\_SIM6Merged\_OLH\_n184\_i6\_o1\_y2.csv} : 184 samples with 6 input and 1 QoI (including the column header titles) provided in CSV format\\
\href{https://mfix.netl.doe.gov/gitlab/quality-assurance/PIC_calibration/-/blob/main/Case1_ParticleSettling/BayesianCalibration/PSUADE/C2_SIM6Merged_OLH_n184_i6_o1_y2.csv}{\tt C2\_SIM6\_OLH\_n120\_i6\_o1\_y2.csv} : 120 samples with 6 input and 1 QoI (including the column header titles) provided in CSV format\\
\href{https://mfix.netl.doe.gov/gitlab/quality-assurance/PIC_calibration/-/blob/main/Case1_ParticleSettling/BayesianCalibration/PSUADE/psData_Merged_n184}{\tt psData\_Merged\_n184} : PSUADE native datafile generated after importing above file CSV or DAT file. Note that only y2:Location of Filling Shock is included as QoI. \\

{\bf Brief Description of the files used: }\\

PSUADE requires an input file with PSUADE command syntax, in addition to an actual dataset, to perform any type of analysis. The default filename for this file is {\tt psuadeData}.  
The reader is strongly advised to use another name, such as {\tt psData}, as PSUADE will overwrite {\tt psuadeData} without warning.

\begin{figure}[H] 
    \begin{center}
    \includegraphics[clip, trim=0.0in 5.12in 0.0in 0.0in, height=0.3\textheight]{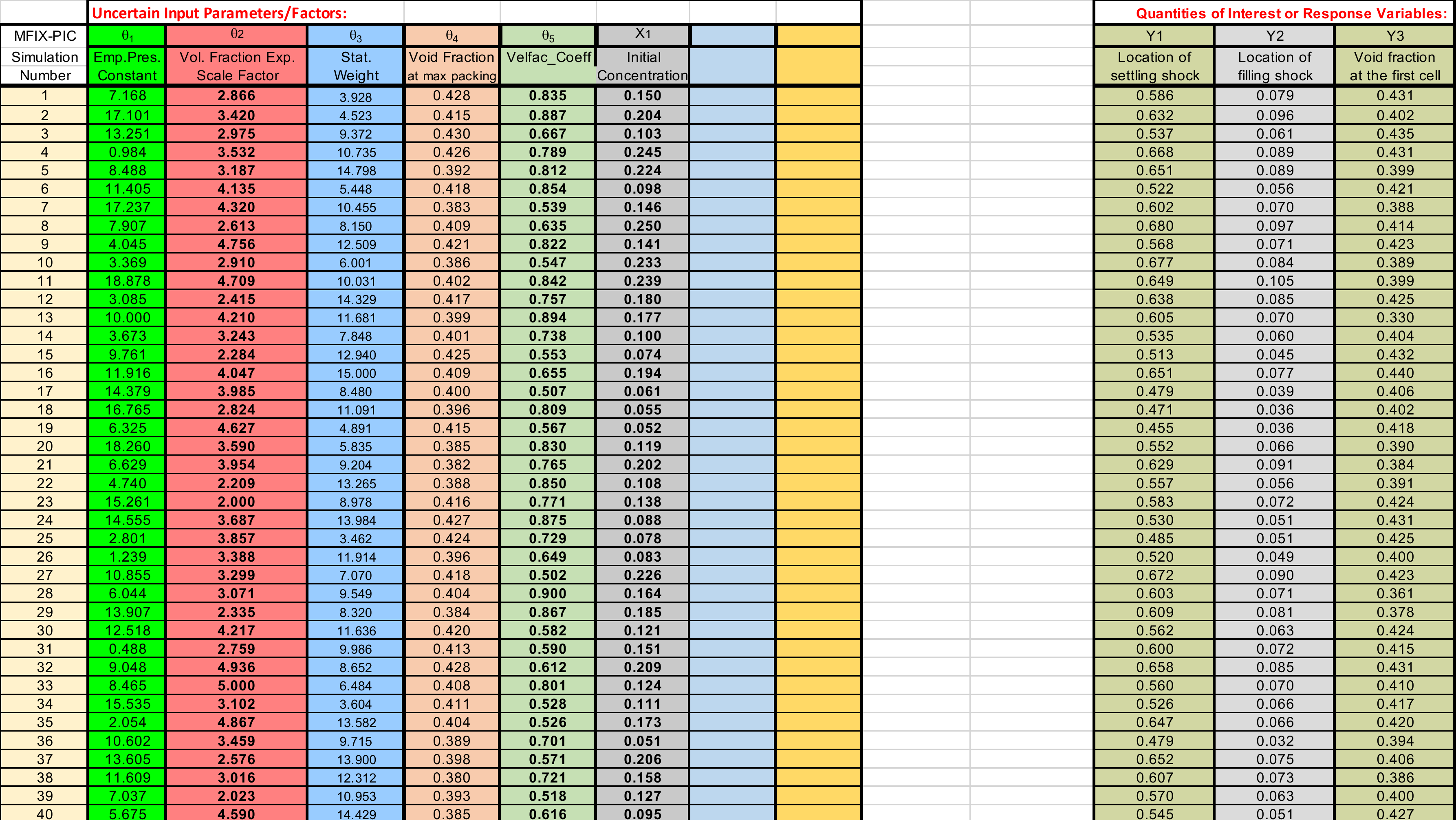}
    \end{center}
    \caption{First 40 samples of the tabulated dataset obtained from simulation campaign \# 1 with 60 samples for six input parameters and three quantities of interest.}
    \label{fig:A_Tabulated_SIM7n60}
\end{figure}

Typically, simulation campaign results are compiled in tabular format.  The first several columns represent input parameters from the design of experiments, and the remaining columns report the associated quantities of interest. Each row represents a single sample from a simulation campaign.  
Figure~\ref{fig:A_Tabulated_SIM7n60} shows a screenshot of the first 40 rows of data from one of the  simulation campaigns in this report.  These were tabulated in Microsoft\textsuperscript{\textregistered} Excel in preparation for analysis in PSUADE. This Microsoft\textsuperscript{\textregistered} Excel
file is saved in the repository under ''{\tt C1\_SIM7\_Results\_OLH\_n60\_i6\_o3.xlsx}'', which is accessible at:\\  \url{https://mfix.netl.doe.gov/gitlab/quality-assurance/PIC_calibration/-/blob/main/Case1_ParticleSettling/DeterministicCalibration_UpdatedResultsDueBugFix/PSUADE/C1_SIM7_Results_OLH_n60_i6_o3.xlsx}.

\begin{figure}[htp]
    \begin{center}
    \includegraphics[clip, trim=0.0in 0.0in 0.0in 0.0in, height=0.3\textheight]{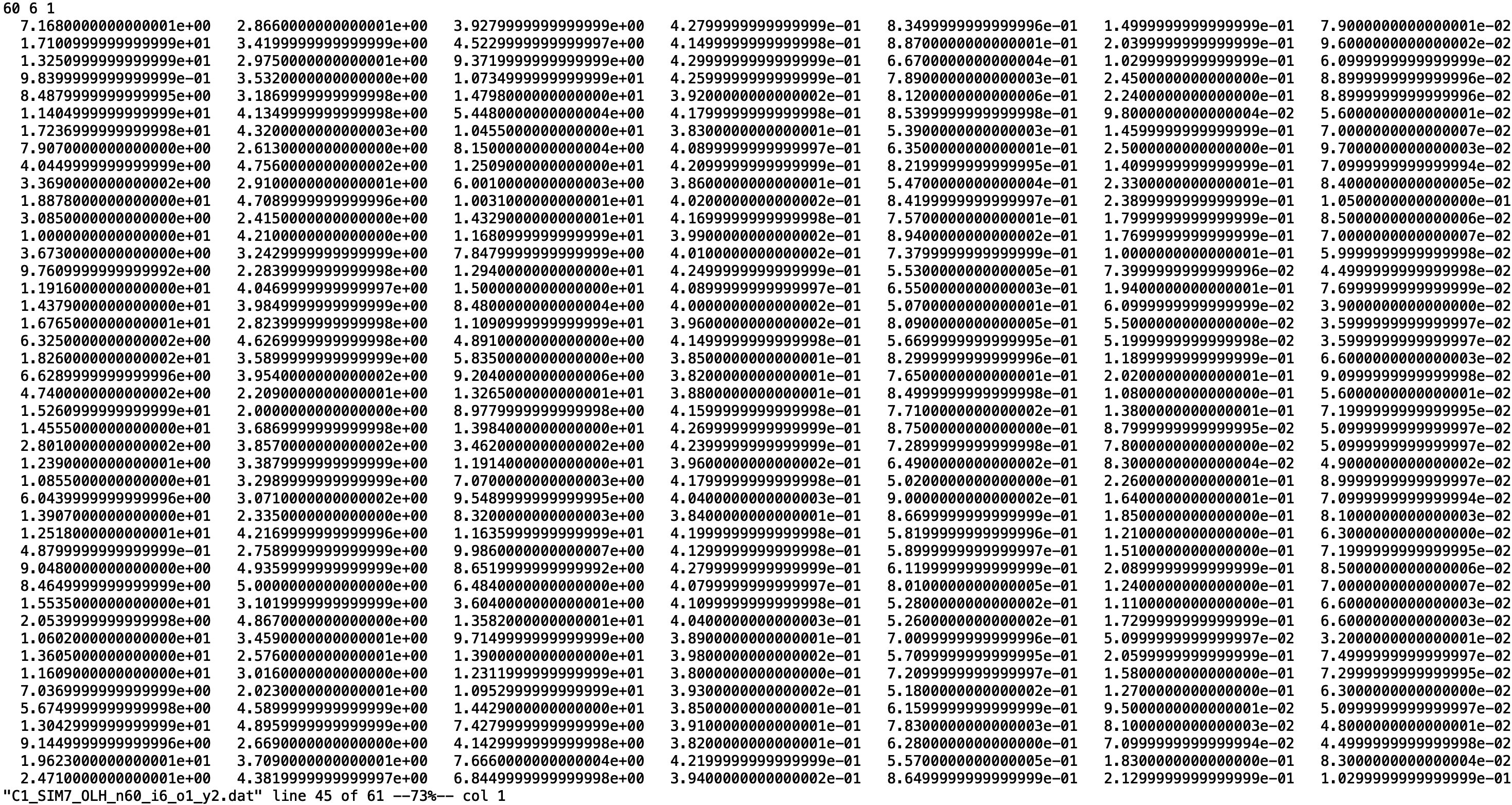}
    \end{center}
    \caption{First 44 lines of the tabulated dataset obtained from simulation campaign \# 1 with 60 samples for six input parameters and one quantity of interest.}
    \label{fig:A_PSTabulated_SIM7n60_o1y2}
\end{figure}

Figure~\ref{fig:A_PSTabulated_SIM7n60_o1y2} shows the first 44 lines of an ASCII formatted file, which is generated from the tabulated results shown in Figure~\ref{fig:A_Tabulated_SIM7n60}. Note that the ASCII file contains an additional header line which indicates that 120 samples of the simulation campaign are included for six input parameters and one output (i.e., only second quantity of interest, $y_2$: Location of filling shock). This file is then imported into PSUADE. This particular ASCII file is saved in the repository under ``{\tt C1\_SIM7\_OLH\_n60\_i6\_o1\_y2.dat},'' which is accessible at: \\
\url{https://mfix.netl.doe.gov/gitlab/quality-assurance/PIC_calibration/-/blob/main/Case1_ParticleSettling/DeterministicCalibration_UpdatedResultsDueBugFix/PSUADE/C1_SIM7_OLH_n60_i6_o1_y2.dat}

The contents of the simulation campaign dataset ({\tt C1\_SIM7\_OLH\_n60\_i6\_o1\_y2.dat}) can be imported into PSUADE with the {\tt read\_std} command while running PSUADE interactively in command line mode. Alternatively, if saved in CSV format {\tt read\_csv} command will import the contents. Details on how to import from an ASCII or CSV file can be found in the PSUADE 1.7 Reference Manual (page 3). 

\begin{figure}[htp]
    \begin{center}
    \includegraphics[clip, trim=0.0in 0.0in 0.0in 0.04in, height=0.4\textheight]{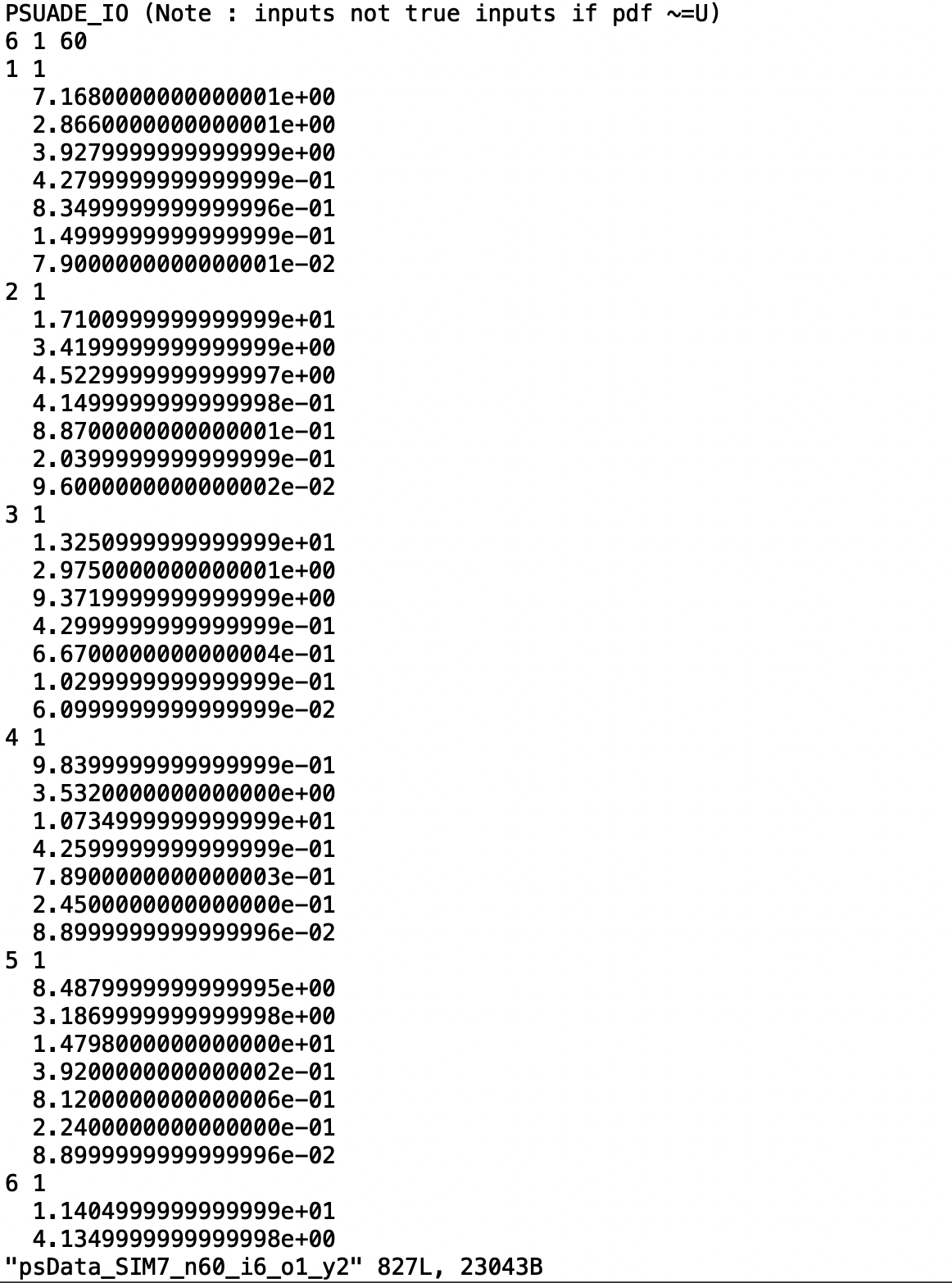}
    \end{center}
    \caption{First 45 lines of PSUADE's native input file {\tt psData\_SIM7\_n60\_i6\_o1\_y2} obtained after importing the dataset in {\tt C1\_SIM7\_OLH\_n60\_i6\_o1\_y2.dat}. }
    \label{fig:A_PS_psData_SIM7n60_o1y2}
\end{figure}

Figure~\ref{fig:A_PS_psData_SIM7n60_o1y2} shows the header segment of the imported file. Note that the ASCII header line used to indicate the number of samples in the simulation campaign, six input parameters and one output, has now been reformatted into a PSUADE native file format. This PSUADE native data file is saved in the repository under "{\tt psData\_SIM7\_n60\_i6\_o1\_y2}" , which is accessible at: \\ \url{https://mfix.netl.doe.gov/gitlab/quality-assurance/PIC_calibration/-/blob/main/Case1_ParticleSettling/DeterministicCalibration_UpdatedResultsDueBugFix/PSUADE/psData_SIM7_n60_i6_o1_y2}.

\begin{figure}[H]
    \begin{center}
    \includegraphics[clip, trim=0.0in 0.04in 0.0in 0.0in, height=0.4\textheight]{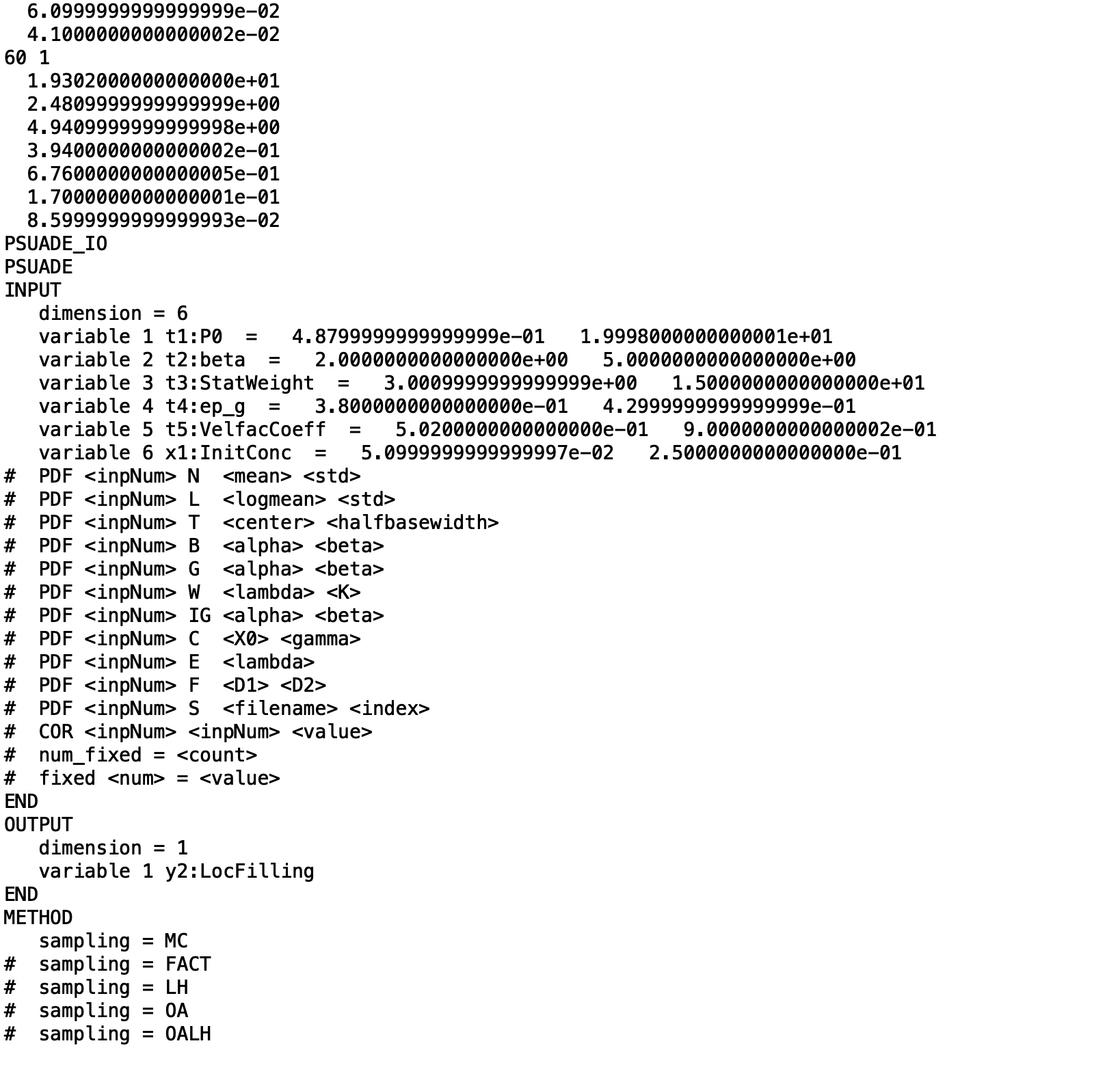}
    \end{center}
    \caption{The INPUT block of PSUADE native datafile ({\tt psData\_SIM7\_n60\_i6\_o1\_y2}  lines: 473-517 shown) which shows the revised labels for input and output parameters.}
    \label{fig:A_PS_psData_INPUT}
\end{figure}

It is important to note that by default PSUADE will generate the file {\tt psData\_SIM7\_n60\_i6\_o1\_y2} with an {\tt INPUT} section showing {\tt x1,x2,x3,$\ldots$} as the name of the input parameters, and {\tt y1,y2, y3,$\ldots$} as the names of the quantities of interest. It is recommended that the user edit {\tt psData\_SIM7\_n60\_i6\_o1\_y2} and rename the input and quantity of interest parameters to indicate their values more appropriately.  For example, in Figure~\ref{fig:A_PS_psData_INPUT}, on line 487, {\tt x1} has been renamed {\tt t1:P\_0}, and on line 510, {\tt y1} has been renamed {\tt y2:LocFilling}). 

As mentioned earlier the files presented in this Appendix and available at the Gitlab repository are the minimum set of files for a researcher with adequate experience in PSUADE to start constructing the surrogate models using the simulation campaign results and then proceed with the follow-up UQ studies such as sensitivity analysis or Bayesian calibration. To avoid confusion for the less experienced user the remaining PSUADE script files were not provided. An inquiry can be submitted to the lead author via email (aike@alpemi.com) for additional information and files.


\end{document}